\documentclass[structabstract]{aa}
\usepackage{float}
\usepackage{lineno}
\usepackage{natbib}
\bibpunct{(}{)}{;}{a}{}{,}
\usepackage{subfigure}
\usepackage{graphicx}
\usepackage{epsfig}
\usepackage{amsmath}
\usepackage[varg]{txfonts}
\usepackage{lscape}
\usepackage{longtable}
\usepackage{txfonts}
\usepackage{color}

\begin{document}
   \title{Nonlinear force-free coronal magnetic field modelling and
        preprocessing of vector magnetograms in spherical geometry}
   \author{Tilaye Tadesse\inst{1,2}, T. Wiegelmann\inst{1}
          \and  B. Inhester\inst{1}
          }
   \institute{Max-Planck-Institut f\"ur Sonnensystemforschung,
              Max-Planck-Strasse 2, 37191 Katlenburg-Lindau, Germany\\
              \email{tadesse@mps.mpg.de; wiegelmann@mps.mpg.de and inhester@mps.mpg.de}
   \and Addis Ababa University, College of Education, Department of Physics
   education, Po.Box 1176, Addis Ababa, Ethiopia}
   \date{Received 19 May 2009 / Accepted 16 September 2009}
  \abstract
     {Knowledge about the coronal magnetic field is important to the understanding of many
    phenomena, such as flares and coronal mass ejections. Routine measurements of the solar magnetic
    field vector are traditionally carried out in the photosphere. We compute the field in the
    higher layers of the solar atmosphere from the measured photospheric field under
    the assumption that the corona is force-free. However, those measured data are inconsistent with the above
    force-free assumption. Therefore, one has to apply some transformations to these data before nonlinear
    force-free extrapolation codes can be applied.}
   {Extrapolation codes of cartesian geometry for medelling the  magnetic field
    in the corona do not take the curvature of the Sun's surface into account.
     Here we develop a method for nonlinear force-free coronal magnetic field medelling and
    preprocessing of photospheric vector magnetograms in spherical geometry
    using the optimization procedure.
  }
   {We describe a newly developed code for the extrapolation of nonlinear force-free
coronal magnetic fields in spherical coordinates over a restricted area of the Sun.
The program uses measured vector magnetograms on the solar photosphere as input and solves the force-free equations in the solar
corona. We develop a preprocessing procedure in spherical geometry to drive the observed non-force-free
   data towards suitable boundary conditions for a force-free extrapolation.}
   {We test the code with the help of a semi-analytic solution and assess the quality of our reconstruction
    qualitatively by magnetic field line plots and quantitatively with a number of comparison metrics for different boundary
    conditions. The reconstructed fields from the lower boundary data with the weighting function are in good agreement with the
   original reference fields. We added artificial noise to the boundary conditions and tested the code with and without
   preprocessing. The preprocessing recovered all main structures of the magnetogram and removed small-scale noise.
   The main test was to extrapolate from the noisy photospheric vector magnetogram with and without preprocessing.
   The preprocessing was found to significantly improve the agreement between the extrapolated and the exact field.}
   {}
\titlerunning{Magnetic field modelling and preprocessing of magnetograms in spherical geometry}
\authorrunning {T. Tadesse \textit{et al}.}
   \maketitle
\section{Introduction}
The magnetic field in the solar corona dominates over non-magnetic forces such as plasma pressure and gravity because of
low plasma beta \citep{Gary}. Knowledge of the coronal magnetic field is therefore important in understanding the
structure of the coronal plasma and obtaining insights into dynamical processes such as flares and coronal mass ejections.
Routine measurements of the solar magnetic field are still mainly carried out in the photosphere.
Therefore, one has to infer the field strength in the higher layers of the solar atmosphere
from the measured photospheric field based on the assumption that the corona is force-free.
 The extrapolation methods involved in this assumption include potential field extrapolation
\citep{Schmidt,Semel67}, linear force-free field extrapolation \citep{Chiu,Seehafer,Seehafer82,Semel88,clegg00}, and
nonlinear force-free field extrapolation
\citep{Amari97,Amari99,Amari,cuperman91,demoulin92,mikic94,Roumeliotis,Sakurai81,valori05,Wheatland04,Wiegelmann04,wu90,yan00}.

 Among these, the nonlinear force-free field has the most realistic
 description of the coronal magnetic field. The computation of nonlinear force-free fields is however, more challenging for
 several reasons. Mathematically, problems regarding the existence and uniqueness of various boundary value problems
dealing with nonlinear force-free fields remain unsolved \citep[see][for details]{Amari}.
Another issue is their numerical analysis of given boundary values.
An additional complication is to derive the boundary data from observed photospheric
vector magnetic field measurements, which are consistent with the force-free
assumption. High noise in the transverse components of the measured
field vector, ambiguities regarding the field direction, and non-magnetic forces in the
photosphere complicate the task of deriving suitable boundary conditions from measured data.
 For a more complete review of existing methods for computing nonlinear force-free coronal
magnetic fields, we refer to the review papers by \citet{Amari97},
\citet{Schrijver06}, \cite{Metcalf}, and \citet{Wiegelmann08}.

 The magnetic field is not force-free in either the photosphere or the lower chromosphere
(with the possible exception of sunspot areas, where the field is strongest). Furthermore,
 measurement errors, in particular for the transverse field components (eg. perpendicular to the line of sight
of the observer), would destroy the compatibility of a magnetogram with the condition of being force-free.
One way to ease these problems is to preprocess the magnetograph data as suggested by \citet{Wiegelmann06sak}.
The vector components of the total magnetic force and the total magnetic torque on the volume considered are given by six
boundary integrals that must vanish if the magnetic field is force-free in the full
volume \citep{Molodenskii69,Aly84,Aly89,Low85}.
 The preprocessing changes the boundary values of  $\vec{B}$ within the error margins of the measurement in such
a way that the moduli of the six boundary integrals are minimized. The resulting boundary values are expected to be more
suitable for an extrapolation into a force-free field than the original values.
In the practical calculations, the convergence properties of the preprocessing iterations, as
well as the calculated fields themselves, are very sensitive to small-scale noise and apparent discontinuities in
the photospheric magnetograph data. This problem should, in principle, disappear if small spatial scales were
sufficiently resolved.
However, the numerical effort for that would be enormous. The small-scale fluctuations in the magnetograms
are also presumed to affect the solutions only in a very thin boundary layer close to the
photosphere \citep{Fuhrmann}. Therefore, smoothing of the data is included in the preprocessing.

The good performance of the optimization method, as indicated in \citet{Schrijver06}, encouraged us to develop a spherical
version of the optimization code such as in \citet{Wiegelmann07} for a full sphere.
In the first few sections of this paper, we describe a newly developed code that originates from a cartesian force-free
optimization method implemented by \citet{Wiegelmann04}. Our new code takes the curvature of the Sun's surface into
account when modeling the coronal magnetic field in restricted area of the Sun. The optimization procedure
considers six boundary faces, but in practice only the bottom boundary face is measured. On the other five faces,
the assumed boundary data may have a strong influence on the solution. For this reason, it is desirable to move these
faces as far away as possible from the region of interest. This, however, eventually requires that the surface
curvature is taken into account.

 \citet{DeRosa} compared several nonlinear force-free codes in cartesian geometry with stereoscopic
reconstructed loops as produced by \citet{Aschwanden}. The codes used as input vector magnetograms
from the Hinode-SOT-SP, which were unfortunately available for only a very small field of view
(about 10 percent of the area spanned by STEREO-loops). Outside the Hinode FOV (field of view) line-of-sight
magnetograms from SOHO/MDI were used and in the MDI-area, different assumptions about the transversal
 magnetic field have been made. Unfortunately, the comparison inferred that when different codes were implemented in the
 region outside the Hinode-FOV in different ways, the resulting coronal magnetic field models produced by the separate codes
 were not consistent with the STEREO-loops. The recommendations of the authors are that
 one needs far larger high resolution vector magnetograms, the codes need to account for uncertainties in the
 magnetograms, and one must have a clearer understanding of the photospheric-to-corona interface. Full disc vector magnetograms
 will soon become available with SDO/HMI, but for a meaningful application we have to take the curvature of the Sun into account
 and carry out nonlinear force-free computations in spherical geometry. In this paper, we carry out the appropriate tests.
 We investigate first ideal model data and later data that contain artificial noise. To deal with noisy data and data
 with other uncertainties, we developed a preprocessing routine in spherical geometry. While preprocessing does not
 model the details of the interface between the forced photosphere and the force-free base of the solar corona the
 procedure helps us to find suitable boundary conditions for a force-free modelling from measurements with inconsistencies.

  In this paper, we develop a spherical version of both the preprocessing and the optimization code for
 restricted part of the Sun. We follow the suggestion of \citet{Wiegelmann06sak} to generalize their method of
 preprocessing photospheric vector magnetograms to spherical geometry just by considering the curvature of the Sun's
 surface for larger field of views.  The paper is organized as follows: in Sect. 2, we describe an optimization procedure
  in spherical geometry; then, in Sect. 3, we apply it to a known nonlinear force-free test field and
  calculate some figures of merit for different boundary conditions. We derive force-free consistency
  criteria and describe the preprocessing procedure in spherical geometry in Sect. 4 and Sect. 5, respectively.
   In Sect. 6, we use a known semi-analytic force-free model to check our method and in Sect. 5, we apply the method
    to different noise models. Finally, in Sect. 7, we draw conclusions and discuss our results.
\section{Optimization procedure}
Stationary states of the magnetic field configuration are described by the requirement that the Lorentz force be
zero. Optimization procedure is one of several methods that have been developed over the past few decades to compute
the most general class of those force-free fields.
\subsection{Optimization principle in spherical geometry }
Force-free magnetic fields must obey the equations
 \begin{equation}
   (\nabla \times\vec{B})\times\vec{B}=0 \,,\label{one}
\end{equation}
\begin{equation}
    \nabla \cdot\vec{B}=0 \label{two}
 \end{equation}
Equations (\ref{one}) and (\ref{two}) can be solved with the help of an optimization principle, as proposed
by \citet{Wheatland00} and generalized by \citet{Wiegelmann04} for cartesian
geometry. The method minimizes a joint measure $(L_\mathrm{\omega})$ of the normalized Lorentz forces and the divergence of the
field throughout the volume of interest, $V$.
Here we define a functional in spherical geometry \citep{Wiegelmann07}:
\begin{equation}
    L_\mathrm{\omega}=\int_{V}\omega(r,\theta,\phi)\Big[B^{-2}\big|(\nabla\times {\vec{B}})\times {\vec{B}}\big|^2+
    \big|\nabla\cdot {\vec{B}}\big|^2\Big]
  r^2\sin\theta dr d\theta d\phi  \,,\label{three}
\end{equation}
where $\omega(r,\theta,\phi)$ is a weighting function and $V$ is a computational
box of wedge-shaped volume, which includes the inner physical domain $V'$ and the
buffer zone(the region outside the physical domain), as shown in Fig. \ref{fig3}
of the bottom boundary on the photosphere. The physical domain $V'$ is a wedge-shaped volume, with two latitudinal boundaries at
$\theta_\mathrm{min}=20^{\degr}$ and $\theta_\mathrm{max}=160^{\degr}$ , two longitudinal boundaries at
$\phi_\mathrm{min}=90^{\degr}$ and
$\phi_\mathrm{max}=270^{\degr}$, and two radial boundaries at the photosphere ($r=1R_{\sun}$) and $r=2R_{\sun}$.
The idea is to define an interior physical region $V'$ in which we wish to calculate the
magnetic field so that it fulfills the force-free or MHS equations. We define $V'$ to be the inner region of $V$
(including the photosphere) with $\omega = 1$ everywhere including its
six inner boundaries $\delta V'$. We use the position-dependent weighting function to introduce a buffer
boundary of $nd = 6$ grid points towards the side and top boundaries of the computational box, $V$.
The weighting function, $\omega$ is chosen to be constant within the inner physical domain $V'$ and declines to 0 with a
cosine profile in the buffer boundary region. The framed region in Figs. \ref{fig3}(a-c) corresponds to the lower
boundary of the physical domain $V'$ with a resolution of $48\times 62$ pixels in the photosphere.

It is obvious that the force-free Eqs. (\ref{one}) and (\ref{two}) are fulfilled when
$L_\mathrm{\omega}$ equals zero.
For fixed boundary conditions, the functional $L_{\omega}$ in Eq. (\ref{three}) can be numerically minimized with
the help of the iteration
\begin{equation}
   \frac{\partial\vec{B}}{\partial t}=\mu\widetilde{\vec{F}} \,,\label{four}
\end{equation}
where $\mu$ is a positive constant and the vector field
$\widetilde{\vec{F}}$ is calculated from
\begin{equation}
   \widetilde{\vec{F}}=\omega \vec{F}+(\Omega_{a}\times\vec{B} )\times\nabla\omega+(\Omega_{b}\cdot\vec{B})
   \nabla \omega \,,\label{five}
\end{equation}
\begin{equation}
   {\vec{F}}=\nabla\times({\vec{\Omega}}_{a}\times {\vec{B}} )-{\vec{\Omega}}_{a}\times(\nabla\times {\vec{B}})+
   \nabla ({\vec{\Omega}}_{b}\cdot {\vec{B}})
   -{\vec{\Omega}}_{b}(\nabla\cdot {\vec{B}})+ \\
   (\Omega_{a}^{2}+\Omega_{b}^{2})\vec{B} \,,\label{six}
\end{equation}
\begin{equation}
   \vec{\Omega}_{a}=\textit{B}^{-2}(\nabla\times\vec{B})\times {\vec{B}} \,,\label{seven}
\end{equation}
\begin{equation}
   \vec{\Omega}_{b}=\textit{B}^{-2}(\nabla\cdot\vec{B})\vec{B} \label{eight}
\end{equation}
The field on the outer boundaries is always fixed here as Dirichlet boundary conditions. Relaxing these
boundaries is possible \citep{Wiegelmann03} and leads to additional terms.
For $\omega (r,\theta,\phi)=1$, the optimization method requires that the
magnetic field is given on all six boundaries of $V'$.
This causes a serious limitation of the method because these data are only
available for model configurations. For the reconstruction of the coronal
magnetic field, it is necessary to develop a method that reconstructs the
magnetic field only from photospheric vector magnetograms \citep{Wiegelmann04}. Since only the bottom
boundary is measured, one has to make assumptions about the lateral and top boundaries, e.g., assume a
potential field. This leads to inconsistent boundary conditions \citep[see][regarding the compatibility of photospheric
vector magnetograph data]{Aly89}. With the help of the weighting function, the five inconsistent boundaries are replaced
by boundary layers and we consequently obtain more flexible boundaries around the physical domain that will be
adjusted automatically during the iteration. This diminishes the effect of the top and lateral boundaries on the
magnetic field solution inside the computational box. Additionally, the influence of the boundaries is diminished,
the farther we move them away from the region of interest.

The theoretical deviation of the iterative Eq. (\ref{four}) as outlined by \citet{Wheatland00}
does not depend on the use of a specific coordinate system. Previous numerical implementations of this method
were demonstrated by \citet{Wiegelmann07} for the full sphere. Within this work, we use a spherical geometry, but for only a
limited part of the sphere, e.g., large active regions, several (magnetically connected) active regions and full
disc computations. Full disc vector magnetograms should become available soon from SDO/HMI.
This kind of computational box will become necessary when the observed photospheric vector
magnetogram becomes available for only parts of the photosphere.

We use a spherical grid $r$, $\theta$, $\phi$ with $n_{r}$, $n_{\theta}$, $n_{\phi}$ grid points in the
direction of radius, latitude, and longitude, respectively. We normalize the magnetic field with the average radial magnetic
field on the photosphere and the length scale with a solar radius.
The method works as follows: \begin{enumerate}
      \item We compute an initial source surface potential field in the computational domain from
$B_{r}$ in the photosphere at $r = 1R_\mathrm{\sun}$.
      \item We replace $B_\mathrm{\theta}$ and $B_{\phi}$ at the bottom photospheric boundary at $r = 1R_{\sun}$ with the measured
vector magnetogram. The outer radial and lateral boundaries are unchanged from the initial potential
field model. For the purpose of code testing, we also tested different boundary conditions (see next section).
 \item We iterate for a force-free magnetic field in the computational box by minimizing the
functional $L$ of Eq.(\ref{three}) by applying Eq.(\ref{four}). For each iteration step ($k$), the vector field
${\widetilde{\vec{F}}}^{(k)}$ is calculated from the known field ${\vec{B}}^{(k)}$, and a new field may simply be computed as
${\vec{B}}^{(k+1)} = {\vec{B}}^{(k)} + {\widetilde{\vec{F}}}^{(k)}\Delta t$ for
sufficiently small $\Delta t$.
  \item The continuous form of Eq.(\ref{four}) ensures a monotonically decreasing functional $L$. For finite
time steps, this is also ensured if the iteration time step $dt$ is sufficiently small. If $L(t +
dt) \geq L(t)$, this step is rejected and we repeat this step with $dt$ reduced by a factor of
$2$.
   \item After each successful iteration step, we increase $dt$ by a factor of $1.01$ to ensure a time
step as large as possible within the stability criteria. This ensures an iteration time step
close to its optimum.
    \item The iteration stops if $dt$ becomes too small. As a stopping criteria, we use $dt \leq10^{-6}$.
   \end{enumerate}
\subsection{Figures of merit}
To quantify the degree of agreement between vector fields $\vec{B}$ (for the input
model field) and $\vec{b}$ (the NLFF model solutions) specified on identical sets
of grid points, we use five metrics that compare either local characteristics (e.g.,
vector magnitudes and directions at each point) or the global energy content in
addition to the force and divergence integrals as defined in \citet{Schrijver06}.
The vector correlation ($C_\mathrm{vec}$) metric generalizes the standard correlation
coefficient for scalar functions given by
\begin{equation}
C_\mathrm{ vec}=  \sum_i \vec{B}_{i} \cdot \vec{ b}_{i}/ \left( \sum_i |\vec{ B}_{i}|^2 \sum_i
|\vec{b}_{i}|^2 \right)^{1/2},\label{nine}
\end{equation}
where $\vec{B}_{i}$ and $\vec{b}_{i}$ are the vectors at each point grid $i$. If the vector fields are identical,
then $C_{vec}=1$; if $\vec{B}_{i}\perp \vec{b}_{i}$ , then
$C_{vec}=0$.\\
The second metric, $C_{CS}$ is based on the Cauchy-Schwarz inequality($|\vec{a}\cdot \vec{b}|\leq
|\vec{a}||\vec{b}|$ for any vector $\vec{a}$ and $\vec{b}$)
\begin{equation}
C_\mathrm{CS} = \frac{1}{N} \sum_i \frac{\vec{B}_{i} \cdot \vec{ b}_{i}} {|{\vec{B}_{i}}||{\vec
{b}_{i}}|},\label{ten}
\end{equation}
where $N$ is the number of vectors in the field. This metric is mostly a measure of the
angular differences between the vector fields: $C_{CS} = 1$, when $\vec{B}$ and $\vec{b}$ are parallel, and
$C_{CS} = -1$, if they are anti-parallel; $C_{CS} = 0$,  if  $\vec{B}_{i}\perp \vec{b}_{i}$ at each
point.\\
We next introduce two measures of the vector errors, one normalized to the
average vector norm, one averaging over relative differences. The normalized vector
error $E_{N}$ is defined as
\begin{equation}
E_\mathrm{N} = \sum_i |{\vec{b}_{i}}-{\vec{B}_{i}}|/ \sum_i |{\vec{B}_{i}}|,\label{eleven}
\end{equation}
The mean vector error $E_{M}$ is defined as
\begin{equation}
E_\mathrm{M} = \frac{1}{N} \sum_i \frac{|{\vec{ b}_{i}}-{\vec{B}_{i}}|}{|{\vec{B}_{i}}|}, \label{twelve}
\end{equation}
Unlike the first two metrics, perfect agreement between the two vector fields results in
$E_{M }= E_{N} = 0$.\\
Since we are also interested in determining how well the models estimate the
energy contained in the field, we use the total magnetic energy in the model field
normalized to the total magnetic energy in the input field as a global measure of
the quality of the fit
\begin{equation}
\epsilon = \frac{\sum_i |{\vec{b}_{i}}|^2}{\sum_i |{\vec{B}_{i}}|^2}, \label{therteen}
\end{equation}
where $\epsilon=1$ for closest agreement between the model field and the nonlinear force-free model
solutions.
\section{Test case and application to ideal boundary conditions}
\subsection{Test case }
To test the method, a known semi-analytic nonlinear solution
is used.  \citet{Low90} presented a class of axisymmetric nonlinear force-free fields with a multipolar
character. The authors solved the Grad-Shafranov equation for
axisymmetric force-free fields in spherical coordinates $r$, $\theta$, and $\phi$. The magnetic field can be
written in the form
\begin{equation}
    {\vec{B}}=\frac{1}{r\sin\theta}\Big( \frac{1}{r}\frac{\partial A}{\partial \theta }{\hat{\bf{e}}}_{r}-
   \frac{\partial A}{\partial r} {\hat{\bf{e}}}_{\theta\ }+Q{\hat{\bf{e}}}_{\phi\ }\Big) \,,\label{fourteen}
\end{equation}
where $A$ is the flux function and $Q$ represents the $\phi$-component of ${\vec{B}}$, depending only on $A$.
The flux function $A$ satisfies the Grad-Shafranov equation
\begin{equation}
   \frac{\partial ^{2} A}{\partial r^{2} }+\frac{1-\mu^{2}}{r^{2}}\frac{\partial ^{2} A}{\partial \mu^{2} }+
   Q\frac{dQ}{dA}=0 \,,\label{fifteen}
 \end{equation}
where $\mu = cos\theta$. \citet{Low90} derive solutions for
\begin{equation}
\\\frac{dQ}{dA}=\alpha=\textrm{const},\label{sixteen}
\end{equation}
by looking for separable solutions of the form
\begin{equation}
\\ A(r,\theta)=\frac{P(\mu)}{r^{n}}\label{seventeen}
\end{equation}
 \citet{Low90} suggested that these field solutions are the ideal solutions for testing methods of reconstructing
force-free fields from boundary values. They have become a standard test for nonlinear force-free extrapolation
codes in cartesian geometry \citep{Amari99,Amari,Wheatland00,Wiegelmann03,Yan06,Inhester06,Schrijver06} because the
symmetry in the solution is no longer obvious after a translation that places the point source
outside the computational domain and a rotation of the symmetry axis with respect to the domain edges.

 Here we use a Low and Lou solution in spherical coordinates. The solution used is labelled
$P_\mathrm{1,1}$ with  $\Phi=\pi/10$, in Low \& Lou's, notation \citep{Low90}. The original equilibrium
is invariant in $\phi$, but we can produce a $\phi$-variation in our coordinate system by placing the origin of
the solution at $l=0.25$ solar radius position from the sun centre. The corresponding configuration
is then no longer symmetric in $\phi$ with respect to the solar surface, as seen in the magnetic field map
in the top row of Fig. \ref{fig3}, which shows the three components $B_{r}$, $B_{\theta}$, and $B_{\phi}$ on the photosphere,
 respectively. We remark that we use the solution only for the purpose of testing our code and the equilibrium is
 not assumed to be a realistic model for the coronal magnetic field. We do the test runs on spherical grids $(r,
\theta,\phi)$ of $20\times 48\times 62$  and $40\times 96\times 124 $ grid points.
\subsection{Application to ideal boundary conditions}
Here we used different boundary conditions extracted from the
 Low and Lou model magnetic field.
\begin{itemize}
\item Case 1: The boundary fields are specified on $V'$(all the six boundaries $\delta V'$ of $V'$).
\item Case 2: The boundary fields are only specified on the photosphere (the lower boundary of
the physical domain $V'$).
\item Case 3: The boundary fields are only specified on the photosphere (the lower boundary of
the physical domain $V'$)  and with boundary layers (at the buffer zone) of $nd=6$  grid points toward top and lateral
boundaries of the computational box $V$.
\end{itemize}

\begin{figure*}[htp!]
   \centering
\mbox{\subfigure[Original ]
   {\includegraphics[bb=120 60 445 350,clip,height=6.0cm,width=7.0cm]{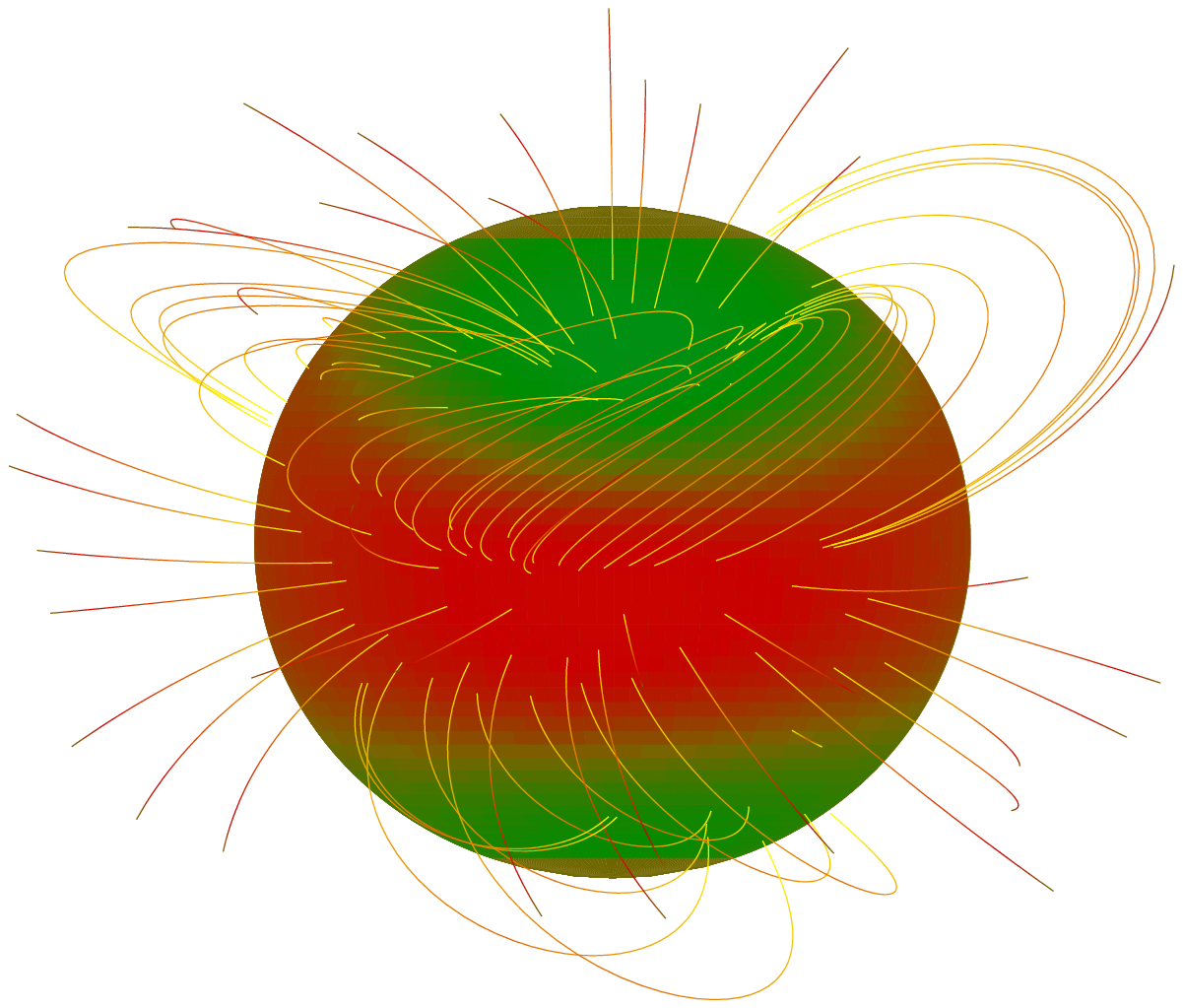}}
\subfigure[Potential]
   {\includegraphics[bb=120 60 445 350,clip,height=6.0cm,width=7.0cm]{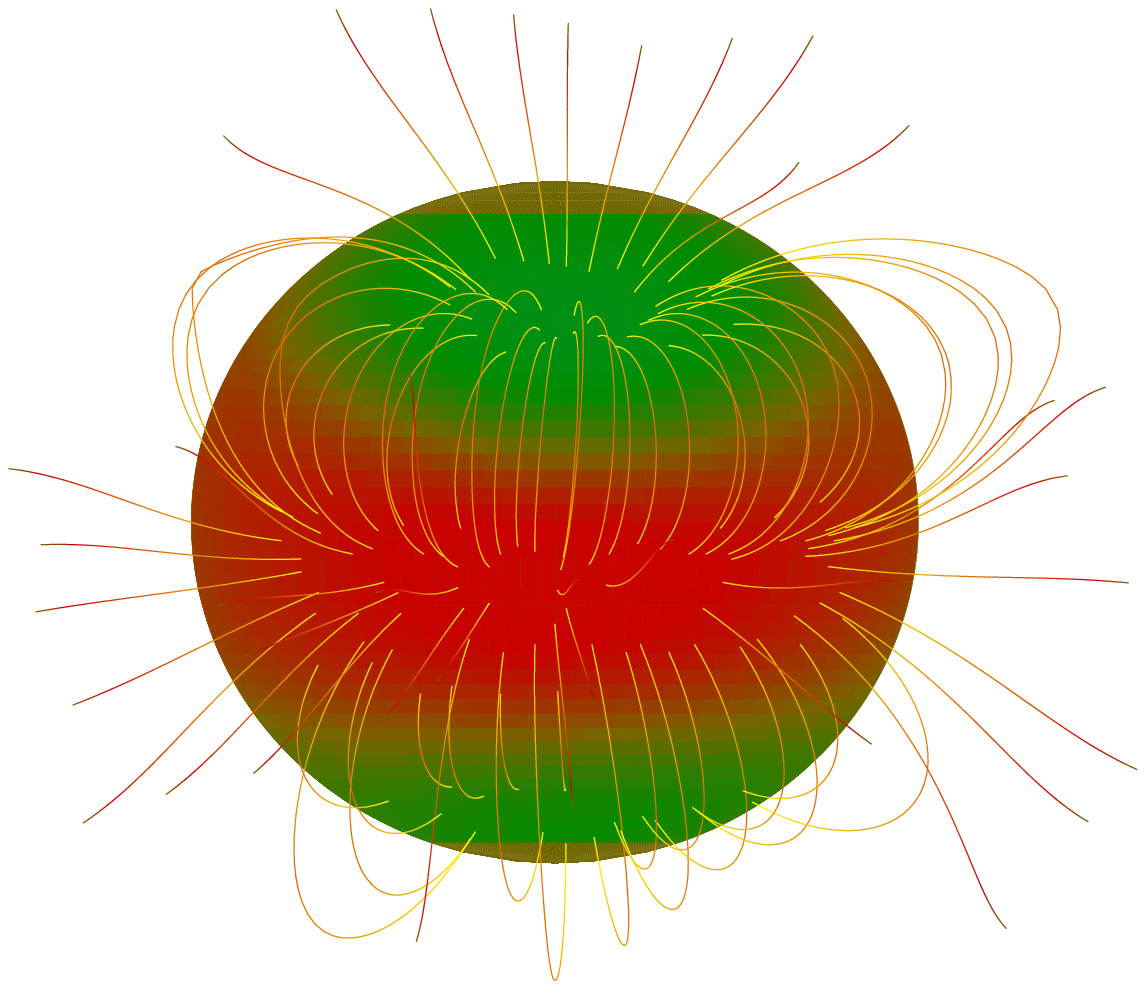}}}
\mbox{\subfigure[Case 1]
   {\includegraphics[bb=120 60 445 350,clip,height=6.0cm,width=7.0cm]{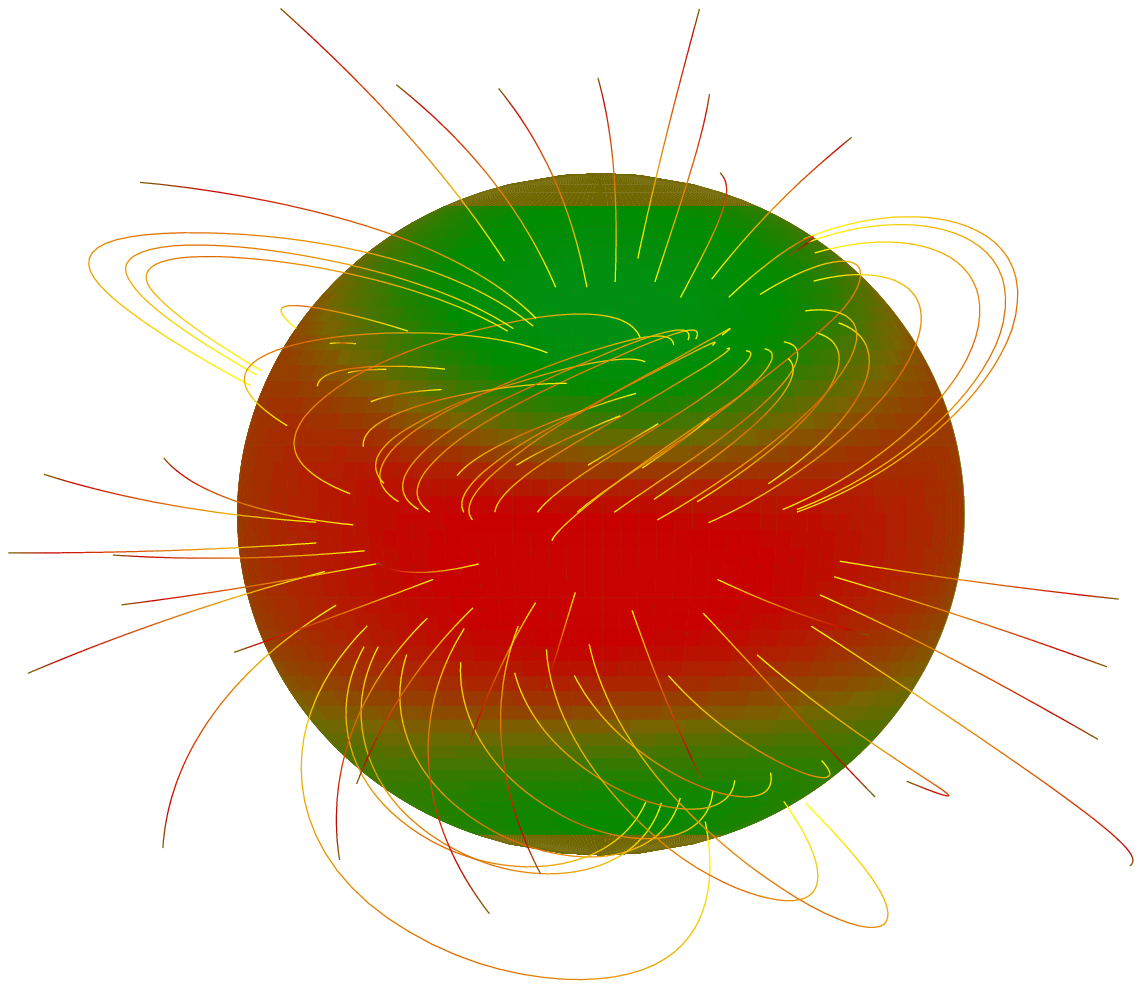}}
\subfigure[Case 2]
   {\includegraphics[bb=120 60 445 350,clip,height=6.0cm,width=7.0cm]{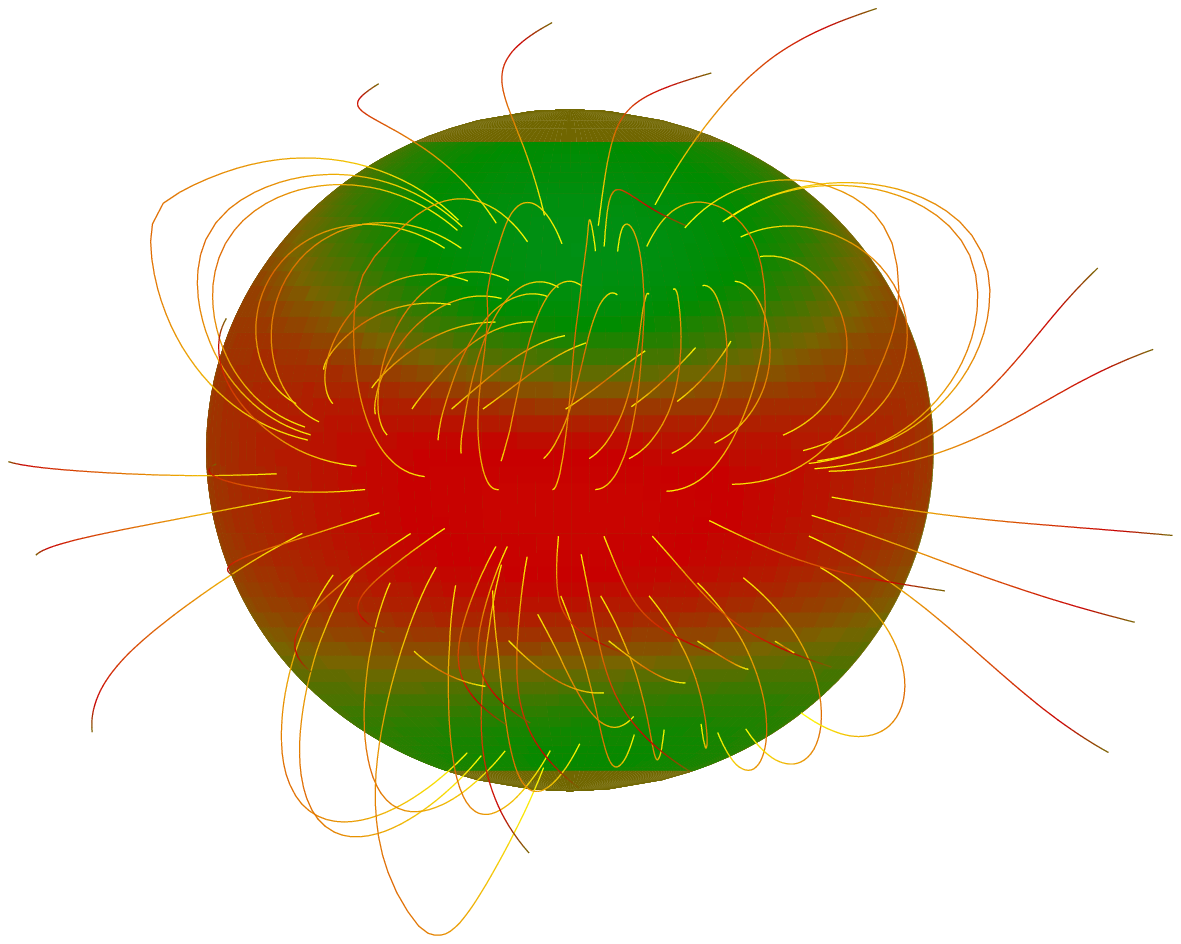}}}
\mbox{\subfigure[Case 3]
   {\includegraphics[bb=120 60 445 350,clip,height=6.0cm,width=7.0cm]{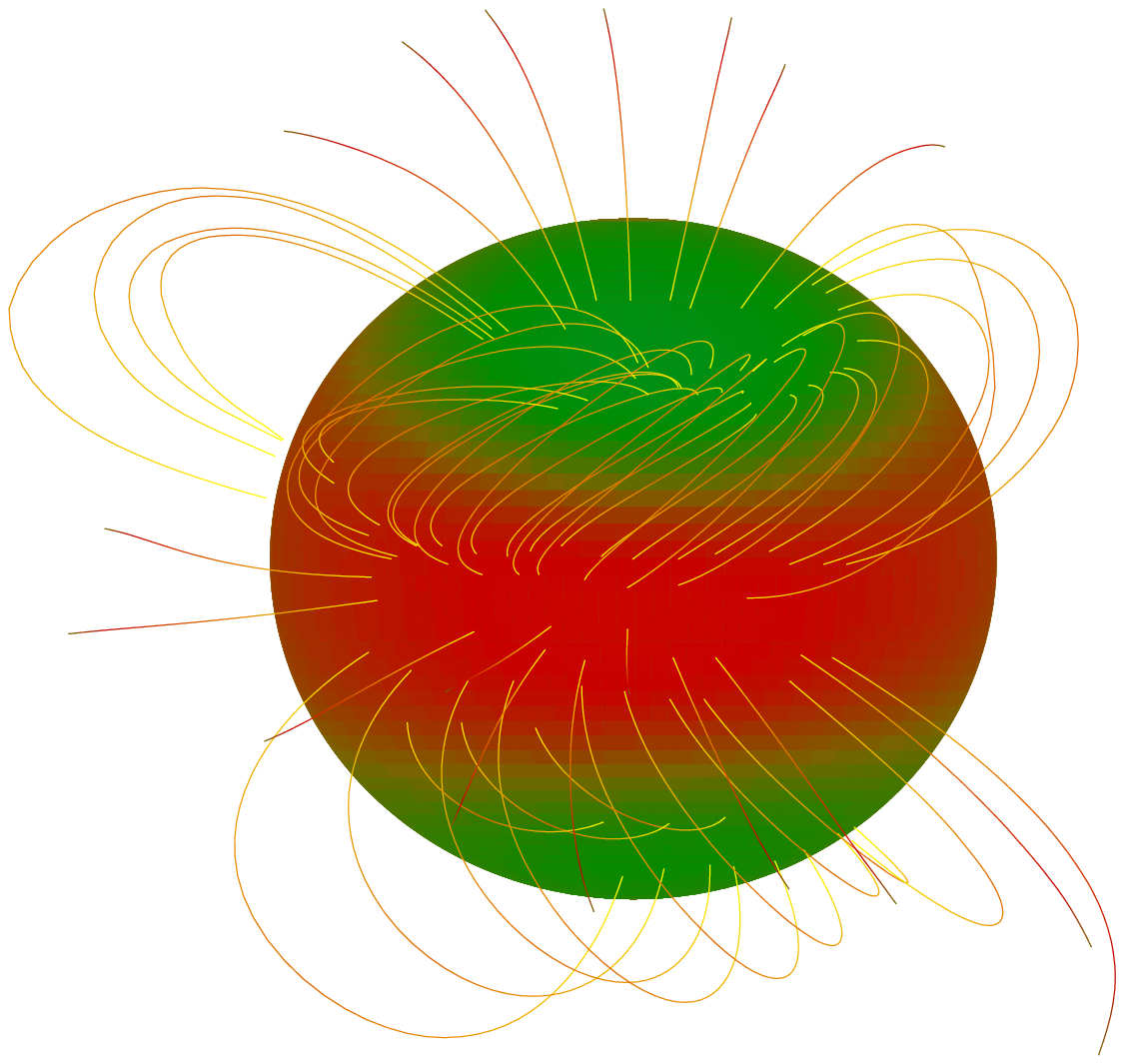}}}
   \caption{The figure shows the original reference field, a  potential field, and
the results of a nonlinear force-free reconstruction with different boundary
conditions (case 1-3, see text). The color coding shows $B_r$ on the photosphere and
the disc centre corresponds to $180^{\degr}$ longitude. }
\label{fig1}
\end{figure*}

For the boundary conditions in case 1, the field line plot (as shown in Fig. $\ref{fig1}$) agrees
with original Low and Lou reference field because the optimization method requires
all boundaries bounding the computational volume as boundary conditions.

For the boundary conditions in case $2$, we used an optimization code without a weighting function ($nd = 0$)
and with a photospheric boundary. Here the boundaries of the physical domain coincide with the computational boundaries.
The lateral and top boundaries assume the value of the potential field during the iteration. Some low-lying
field lines are represented quite well (right-hand picture in Fig. $\ref{fig1}$ second row).
The field lines close to the box center are of course close to the bottom boundary
and far away from the other boundaries. The (observed) bottom boundary has a
higher influence on the field here than the potential lateral and top boundary. Other
field lines, especially high-reaching field lines, deviate from the analytic solution.

  For the boundary condition in case $3$, we implemented an optimization code with a weighting function of $nd = 6$
  grid points outside the physical domain. This reduces the effect of top and lateral boundaries where $\vec{B}$ is
  unknown as $\omega$ drops from $1$ to $0$ outward across the boundary layer around the physical domain.

 The comparison of the field lines of the Low \& Lou model field with the reconstructed field of case $3$
 (the last picture in Fig. \ref{fig1})  shows that the quality of the reconstruction
improves significantly with the use of the weighting function. Additionally, the size and shape of a boundary layer
influences the quality of the reconstruction \citep{Wiegelmann04} for cartesian geometry.
The larger computational box displaces the lateral and top boundary further away from the physical domain and
 its influence on the solution consequently decreases. As a result, the magnetic field in the physical domain is
dominated by the vector magnetogram data, which is exactly what is required for application to measured vector magnetograms.
 A potential field reconstruction obviously does not agree with the reference field. In particular, we are in able to
 compute the magnetic energy content of the coronal magnetic field to be approximately correct. The figures of merit show
 that the potential field is far away from the true solutions and contains only $67.6\%$ of the magnetic energy.
\begin{table*}
\caption {
Quality of our reconstructions with several figures of merit as explained
in section 3.2. We compute the figures for the three different cases along with the model reference field and potential
field.}
\hspace*{-0.5cm}
\centering
\begin{tabular}{l|lllcc|lllll|rr}     
\hline \hline Model & $L_{\omega}$&$L_{f}$&$L_{d}$&
$\parallel \nabla \cdot {\vec{B}} \parallel_{\infty}$&
$\parallel {\vec{ j} } \times {\vec{B}} \parallel_{\infty} $ &
$C_{\rm vec}$&$C_{\rm CS}$&$E_{N}$&$E_{M}$&$\epsilon$ &Steps& Time \\
\hline
&\multicolumn{3}{c}{Spherical grid $20 \times 48 \times 62$} &&&&&&&& \\
Original &$0.029$&$0.015$&$0.014$&$1.180$&$1.355$& $ 1$&$ 1$&$ 0$&$ 0$&$ 1$&$ $&$ $ \\
Potential &$0.020$&$0.007$&$0.014$&$1.706$&$1.091$&$0.736$&$0.688$&$0.573$&$0.535$&$0.676$&& \\
Case 1 &$0.006$&$0.004$&$0.002$&$0.454$&$0.774$&$0.999$&$0.983$&$0.012$&$0.016$&$1.005$&$10000$& 7.14 min \\
Case 2 &$33.236$&$7.806$&$25.430$&$47.843$&$24.135$&$0.757$&$0.726$&$0.397$&$0.451$&$0.745$&$110$& 1.28 min \\
Case 3 &$0.009$&$0.006$&$0.03$&$0.367$&$0.787$&$0.994$&$0.967$&$0.187$&$0.097$&$0.989$&$12011$& 17.54 min \\
\hline
&\multicolumn{3}{c}{Spherical grid $40 \times 96 \times 124$} &&&&&&&& \\
Original &$0.005$&$0.003$&$0.002$&$0.38$&$0.71$&$ 1$&$ 1$&$ 0$&$ 0$&$ 1$&$ $&$ $ \\
Potential &$0.30$&$0.0003$&$0.30$&$0.44$&$0.23$&$0.67$&$0.77$&$0.70$&$0.67$&$0.75$&$ $&$ $ \\
Case 1 &$0.002$&$0.001$&$0.0006$&$0.38$&$0.32$&$0.998$&$0.999$&$0.004$&$0.007$&$1.001$&$12522$& 1h 21min \\
Case 2 &$26.27$&$10.20$&$16.07$&$20.40$&$30.53$&$0.799$&$0.759$&$0.411$&$0.456$&$0.798$&$5673$&1h 1min \\
Case 3 &$0.24$&$0.20$&$0.04$&$0.630$&$0.747$&$0.996$&$0.971$&$0.186$&$0.112$&$0.996$&$12143$ & 4h 57min\\
\hline
\end{tabular}
\end{table*}

\begin{figure}
   \centering
   \includegraphics[bb=55 375 552 693,clip,height=6.0cm,width=8.5cm]{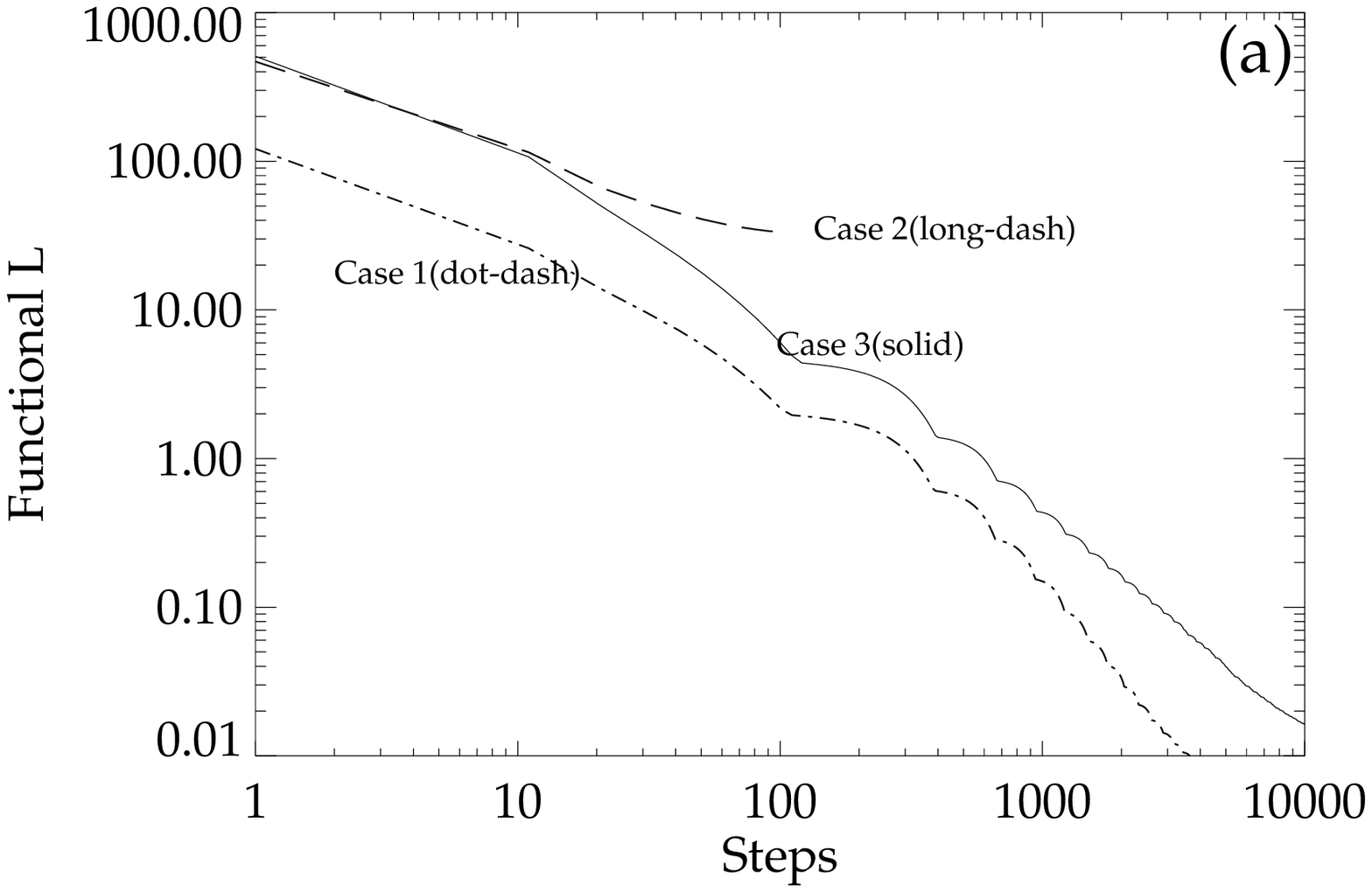}
   \includegraphics[bb=55 375 552 693,clip,height=6.0cm,width=8.5cm]{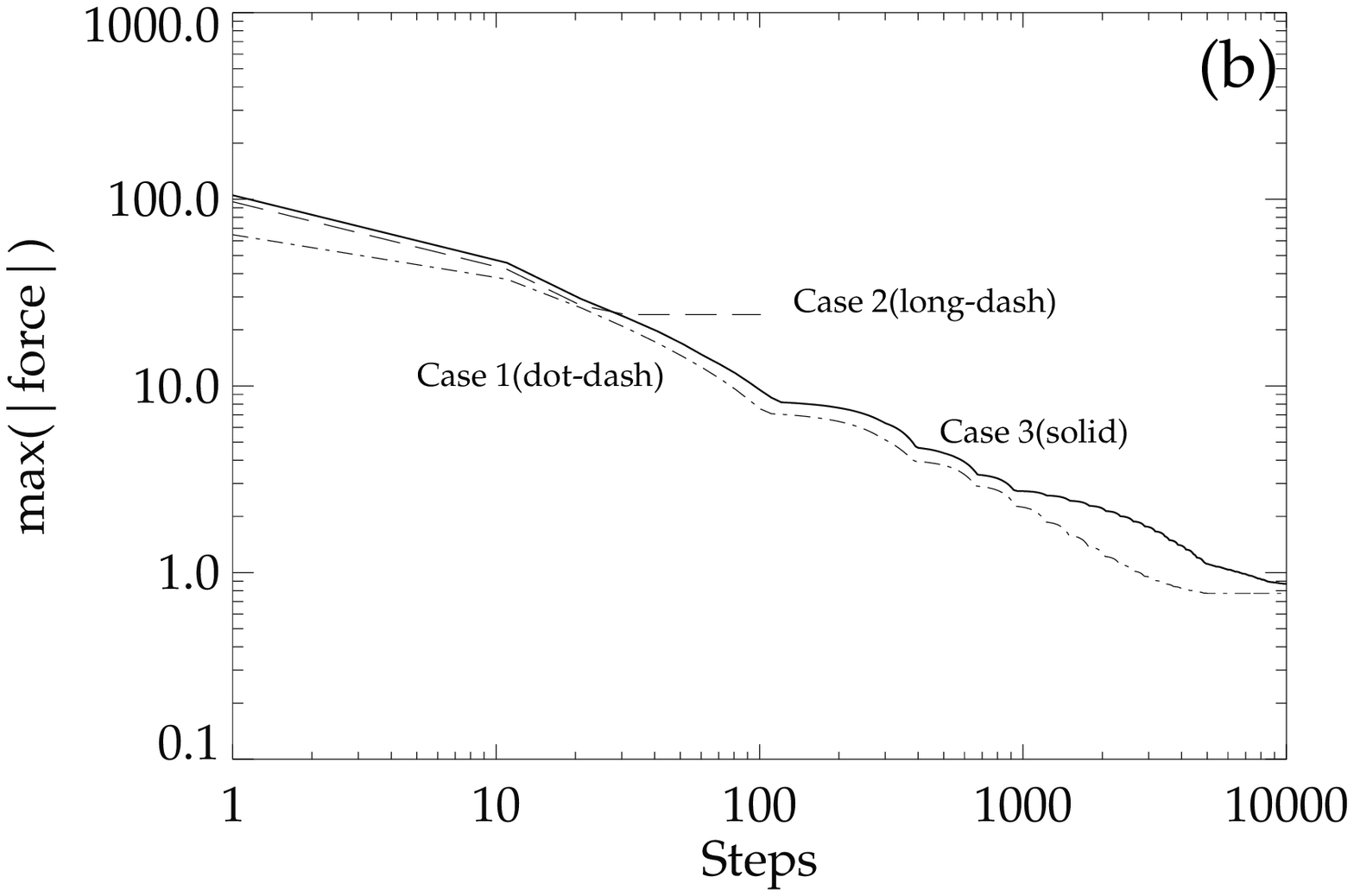}
    \includegraphics[bb=55 375 552 693,clip,height=6.0cm,width=8.5cm]{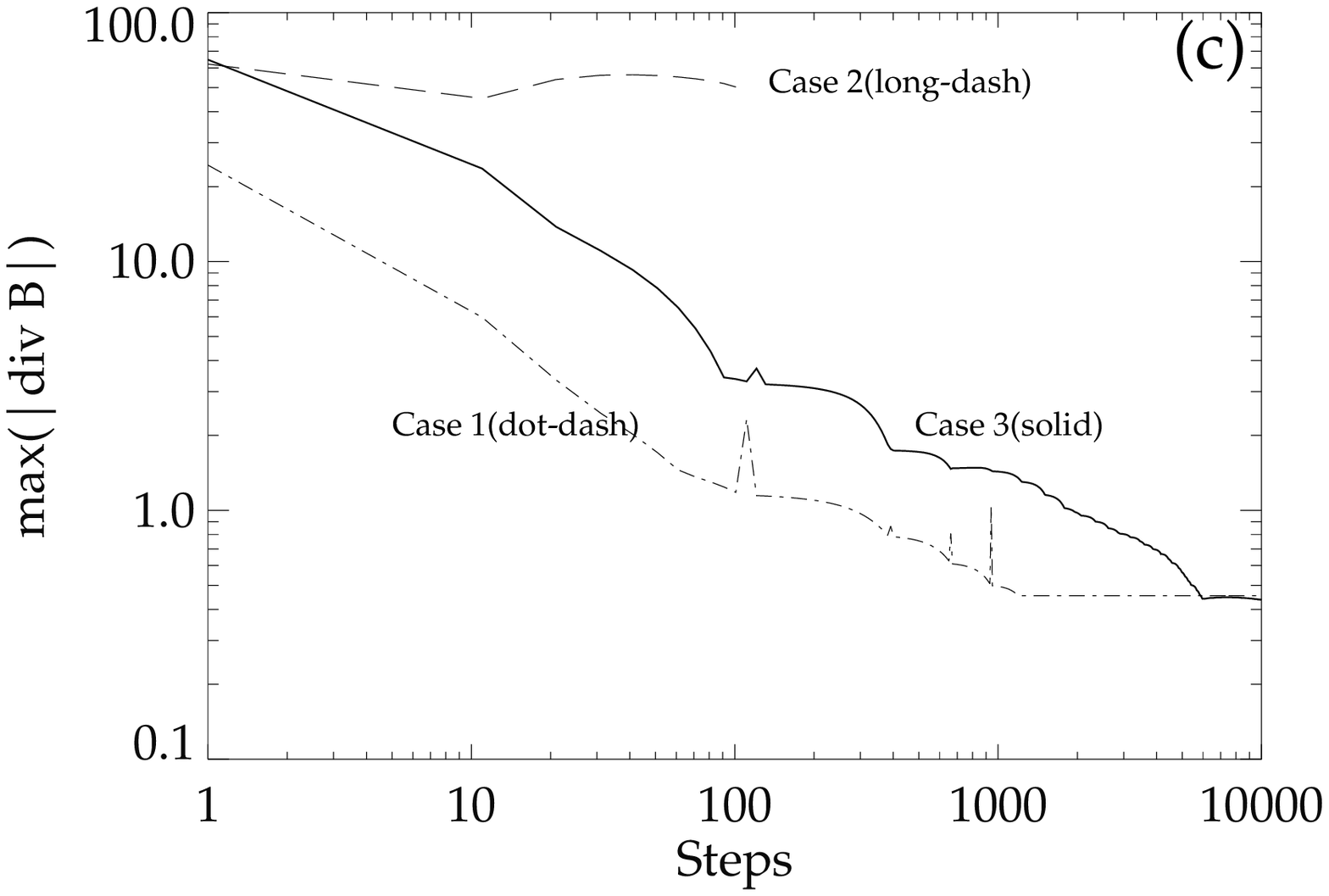}
       \caption{Evolution of $L_{\omega}$ (as defined in Eq. 3), $max(|force|)$, and $max(|div B|)$ during the optimization process.
       The solid line corresponds to case 3, the dash-dotted line to case 1, the long-dashed line to case 2.}
\label{fig2}
       \end{figure}

The degree of convergence towards a force-free and divergence-free model solution
can be quantified by the integral measures of the Lorentz force and divergence
terms in the minimization functional in Eq. (\ref{three}), computed over the entire
model volume $V$:
\begin{displaymath} L_{f}=\int_{V}\omega(r,\theta,\phi)B^{-2}\big|(\nabla\times {\vec{B}})\times
{\vec{B}}\big|^2  r^2\sin\theta dr d\theta d\phi ,
\end{displaymath}
\begin{displaymath}L_{d}=\int_{V}\omega(r,\theta,\phi)\big|\nabla\cdot {\vec{B}}\big|^2
  r^2\sin\theta dr d\theta d\phi ,
\end{displaymath}
\begin{displaymath}L_{\omega}=L_{f}+L_{d},
\end{displaymath}
  where $L_{f}$ and $L_{d}$ measure how well the force-free and divergence-free conditions
are fulfilled, respectively.
In Table $1$, we list the figures of merit for our extrapolations results as
introduced in previous section.
Column 1 indicates the corresponding test case. Columns $2-4$ show how well the force and
solenoidal condition are fulfilled, where Col. $2$ contains the value of the functional
$L_{\omega}$
as defined in Eq.$(3)$ and $L_{f}$ and $L_{d}$ in Cols. $3$ and $4$ correspond to the first (force-free) and
second (solenoidal free) part of $L_{\omega}$. The evolution of the functional $L_{\omega}$,  $|\vec{j}\times \vec{B}|$,
 and   $|\nabla \cdot \vec{B}|$ during the optimization process is shown in Fig. \ref{fig2}.  One can see from this figure that
the calculation does not converge for case 2, because of the problematic boundaries where the fields are unknown.
 Column $5$ contains the $L^{\infty}$ norm of the divergence of the magnetic field
\begin{displaymath}\parallel \nabla \cdot {\vec{ B}} \parallel_{\infty}=\sup_{{\bf x} \in V} |\nabla \cdot {\vec{B}}|
\end{displaymath}
and Col. $6$ lists the $L^{\infty}$ norm of the Lorentz force of the magnetic field
\begin{displaymath}\parallel {\vec{j} } \times {\vec{B}} \parallel_{\infty}=\sup_{{\bf x} \in V} |{\vec{j} } \times {\vec{B}} |.
\end{displaymath}
The next five columns of Table $1$ contain different measurements comparing our reconstructed field with the semi-analytic
reference field. The two vector fields agree perfectly if $C_{\rm vec}$, $C_{\rm CS}$, and $\epsilon$ are
unity and if $E_{\rm N}$ and $E_{\rm M}$ are zero. Column 12 contains the number of
iteration steps until convergence, and Col. 13 shows the computing time on $1$ processor.

 A comparison of the original reference field (Fig. \ref{fig1}(a)) with our nonlinear force-free reconstructions
 (cases 1-3) shows that the magnetic field line plots agree with the original for case 1 and case 3
 within the plotting precision. Case 2 shows some deviations from the original, but the reconstructed field lines are much
 closer to the reference field than the initial potential field.
The visual inspection of Fig. \ref{fig1} is supported by the quantitative criteria shown in Table 1. For case 1 and case 3
the formal force-free criteria $(L_{\omega},L_{f},L_{d})$ are smaller than the discretization error of the analytic
solution and the comparison metrics show almost perfect agreement with the reference field. The comparison metrics
(of Table 1) show that there is discrepancy between the reference field and case 2 as the magnetic field solution is
affected by nearby problematic top and lateral boundaries. In Fig. \ref{fig1} we compare magnetic field line plots of
 the original model field with a corresponding potential field and nonlinear force-free reconstructions with different
 boundary conditions (case 1 - case 3). The colour coding shows the radial magnetic field in the photosphere, as also shown in the
 magnetogram in Fig. $\ref{fig3}$(a).  The images show the results of the computation on the $20 \times 48 \times 62$ grid.
\section{Consistency criteria in spherical geometry}
A more fundamental requirement of the boundary data is its consistency with the force-free field approximation. As
shown by  \citet{Molodenskii69} and \citet{Aly89}, a balance between the total momentum and angular momentum
exerted onto the numerical box in cartesian geometry by the magnetic field leads to a set of
boundary integral constraints on the magnetic field. These constraints should also be satisfied on the solar surface
for the field at the coronal base in the vicinity of a sufficiently isolated magnetic region and in a
situation where there is no rapid dynamical development. As explained in detail in
\citet{Molodenskii74}, the sense of these relations is that on average a force-free field cannot exert a net tangential
force on the boundary or shear stresses along axes lying along the boundary.
In summary, the boundary data for the force-free extrapolation should fulfill the following conditions:
\begin{enumerate}
\item The boundary data should coincide with the photospheric observations within
measurement errors.
\item The boundary data should be consistent with the assumption of a force-free
magnetic field above.
\item For computational reasons (finite differences), the boundary data should be
sufficiently smooth.
\end{enumerate}
Additional a-priori assumption is about the photospheric data are that the magnetic flux from the photosphere is
sufficiently distant from the boundaries of the observational domain and that the net flux is balanced,
i.e.,
\begin{equation}
\int_{S}B_{r}(r=1R_{s},\theta,\phi)d\Omega =0,\label{eighteen}
\end{equation}
where $S$ is the area of a bottom boundary of the physical domain on the photosphere.

Generally, the flux balance criterion must be applied to the entire, closed surface of the numerical box. However, we can
only measure the magnetic field vector on the bottom photospheric boundary and the contributions of the lateral and top
boundary remain unspecified. However, if a major part of the known flux from the bottom boundary
is uncompensated, the final force-free magnetic field solution will depend markedly on how the uncompensated flux is
distributed over the other five boundaries. This would result in a major uncertainty on the final force free magnetic
field configuration. We therefore demand that the flux balance is satisfied with the bottom
data alone \citep{Wiegelmann06}. If this is not the case, we classify the reconstruction problem as not being uniquely
solvable within the given box. \citet{Aly89} used the virial theorem to define the conditions that a
vector magnetogram must fulfill to be consistent with the assumption of a force-free field above in cartesian geometry.
Here in this paper, we assume the force-free and torque-free conditions for spherical geometry as formulated in \citet{Sakurai94},
i.e.,

1. The total force on the boundary vanishes
\begin{equation}
\int_{S}\Big[\frac{1}{2}\big(B_{\theta}^{2}+B_{\phi}^{2}-B_{r}^{2}\big)\sin\theta\cos\phi-B_{r}B_{\theta}\cos\theta\cos\phi
 + B_{r}B_{\phi}\sin\phi\Big]d\Omega=0,\label{nineteen}
\end{equation}
\begin{equation}
\int_{S}\Big[\frac{1}{2}\big(B_{\theta}^{2}+B_{\phi}^{2}-B_{r}^{2}\big)\sin\theta\sin\phi-B_{r}B_{\theta}\cos\theta\sin\phi
-B_{r}B_{\phi}\cos\phi\Big]d\Omega=0,\label{twenty}
\end{equation}
\begin{equation}
\int_{S}\Big[\frac{1}{2}\big(B_{\theta}^{2}+B_{\phi}^{2}-B_{r}^{2}\big)\cos\theta+B_{r}B_{\theta}\sin\theta\Big]d\Omega=0
\label{twenty_one}
\end{equation}

2. The total torque on the boundary vanishes\footnote{See Appendix A for derivation of those torque-balance equations.}
\begin{equation}
\int_{S}B_{r}\big(B_{\phi}\cos\theta\cos\phi+B_{\theta}\sin\phi\big)d\Omega=0,\label{twenty_two}
\end{equation}
\begin{equation}
\int_{S}B_{r}\big(B_{\phi}\cos\theta\sin\phi-B_{\theta}\cos\phi\big)d\Omega=0,\label{twenty_three}
\end{equation}
\begin{equation}
\int_{S}B_{r}B_{\phi}\sin\theta d\Omega=0\label{twenty_four}
\end{equation}
As with the flux balance, these criteria must in general, be applied to the entire surface of the numerical box.
Since we assumed that the photospheric flux is sufficiently concentrated in the center and the net flux is in balance,
we can expect the magnetic field on the lateral and top boundaries to remain weak and hence these surfaces do not
represent a significant contribution to the integrals of the constraints above. We therefore impose the criteria on
the bottom boundary alone. From this beginning, we use the following notation for simplicity:
\begin{displaymath}
      E_{B}^{-}=\frac{1}{2}\big(B_{\theta}^{2}+B_{\phi}^{2}-B_{r}^{2}\big)    \, , \;
E_{B}=\int_{S}\big(B^{2}_{r}+B^{2}_{\theta}+B^{2}_{\phi}\big)d\Omega
\, , \;
\end{displaymath}
\begin{displaymath}
B_{1}=B_{\theta}\cos\theta\cos\phi-B_{\phi}\sin\phi   \, , \;
B_{2}=B_{\theta}\cos\theta\sin\phi+B_{\phi}\cos\phi
\, , \;
\end{displaymath}
\begin{displaymath}
B_{3}=B_{\phi}\cos\theta\cos\phi+B_{\theta}\sin\phi       \, , \;
B_{4}=B_{\phi}\cos\theta\sin\phi-B_{\theta}\cos\phi
 \;
\end{displaymath}
To quantify the quality of the vector magnetograms with respect to the above criteria, we introduce three dimensionless
parameters similar to those in \citet{Wiegelmann06sak}, but now for spherical geometry:
\begin{enumerate}
\item The flux balance parameter
$$\varepsilon_{flux}=\frac{\int_{S}B_{r}d\Omega}{\int_{S}|B_{r}|d\Omega}$$
\item The force balance parameter
\begin{displaymath}\begin{split}
\varepsilon_{force}=&\Big(\big| \int_{S}\big[E_{B}^{-}\sin\theta\cos\phi
 -B_{r}B_{1}\big]d\Omega\big|
 +\big|\int_{S}\big[E_{B}^{-}\sin\theta\sin\phi
 \\ &-B_{r}B_{2}\big]d\Omega\big|
+\big|\int_{S}\big[E_{B}^{-}\cos\theta  +B_{r}B_{\theta}\sin\theta\big]d\Omega\big|\Big)
  \Big/E_{B}
\end{split}\end{displaymath}
\item The torque balance parameter
\begin{displaymath}\begin{split}
\varepsilon_{torque}=&\Big(\big|\int_{S}B_{r}B_{3}d\Omega\big|+\big|\int_{S}B_{r}B_{4}d\Omega\big|
+\big|\int_{S}B_{r}B_{\phi}\sin\theta d\Omega\big|\Big)\Big/E_{B}
\end{split}\end{displaymath}
\end{enumerate}
An observed vector magnetogram is then flux-balanced and consistent
with the force-free assumption if:  $\varepsilon_{flux}\ll 1$, $\varepsilon_{force}\ll
1$ and $\varepsilon_{torque}\ll 1$.
\section{Preprocessing method}
The strategy of preprocessing is to define a functional $L$ of the boundary values of $\vec{B}$, such that on
minimizing $L$ the total magnetic force and the total magnetic torque on the considered volume,
as well as a quantity measuring the degree of small-scale noise in the boundary data, simultaneously become small. Each
of the quantities to be made small is measured by an appropriately defined subfunctional included in $L$. The different
subfunctionals are weighted to control their relative importance.
Even if we choose a sufficiently flux balanced isolated active region ($\varepsilon_{flux}\ll 1$), we find that the force-free
conditions $\varepsilon_{force}\ll 1$ and $\varepsilon_{torque}\ll 1$ are not usually fulfilled for measured
vector magnetograms. We therefore conclude, that force-free extrapolation methods
should not be used directly on observed vector magnetograms (see \citet{Gary} for $\beta >1$ in photosphere), particularly
 not on very noisy transverse photospheric magnetic field measurements. The large noise in the transverse
components of the photospheric field vector, which is one order of magnitude higher than the LOS-field
 ($\sim$the transverse $B_{\theta}$ and $B_{\phi}$ at the bottom boundary), provides us freedom to adjust these data within the
noise level. We use this freedom to drive the data towards being more consistent with Aly's force-free and torque-free conditions.

The preprocessing scheme of \citet{Wiegelmann06sak} involves
 minimizing a two-dimensional functional of quadratic form similar to the following:
\begin{equation}
L=\mu_{1}L_{1}+\mu_{2}L_{2}+\mu_{3}L_{3}+\mu_{4}L_{4}\label{twenty_five}
\end{equation}
Here we write the individual terms in spherical co-ordinates as: \begin{equation}\begin{split}
L_{1}= \Big(\sum_{p}\big[E_{B}^{-}\sin\theta\cos\phi
-B_{r}B_{1}\big]\sin\theta\Big)^{2}
+\Big(\sum_{p}\big[ E_{B}^{-}\sin\theta\sin\phi
\\-B_{r}B_{2}\big]\sin\theta\Big)^{2}
+ \Big(\sum_{p}\Big[E_{B}^{-}\cos\theta
+B_{r}B_{\theta}\sin\theta\big]\sin\theta\Big)^{2},\label{twenty_six}
\end{split}\end{equation}
\begin{equation}
L_{2}=\Big(\sum_{p} B_{r}B_{3}\sin\theta\Big)^{2}
+\Big(\sum_{p}B_{r}B_{4}\sin\theta \Big)^{2}
+\Big(\sum_{p} B_{r}B_{\phi}\sin^{2}\theta\Big)^{2},\label{twenty_seven}
\end{equation}
\begin{equation}
 L_{3}=\sum_{p}\big(B_{r}-B_{robs}\big)^{2}+\sum_{p}\big(B_{\theta}-B_{\theta obs}\big)^{2}
 +\sum_{p}\big(B_{\phi}-B_{\phi obs}\big)^{2},\label{twenty_eight}
\end{equation}
\begin{equation}
L_{4}=\sum_{p}\big[ \big(\Delta B_{r}\big)^{2}+\big(\Delta B_{\theta}\big)^{2}
+\big(\Delta B_{\phi}\big)^{2}\big]\label{twenty_nine}
\end{equation}
The surface integrals are replaced by a summation $\Big(\int_{S}d\Omega\rightarrow\Sigma_{p}
\sin\theta\Delta\theta\Delta\phi$, omitting the constant $\Delta\theta\Delta\phi$ over all
$p$ grid nodes of the bottom surface grid, with an elementary surface of $\sin\theta\Delta\phi\times\Delta\theta$\Big).
The differentiation in the smoothing term ($L_{4}$) is achieved by the usual five-point
stencil for the 2D-Laplace operator. Each of the constraints $L_{n}$ is weighted by a yet undetermined factor
$\mu_{n}$. The first term $(n=1)$ corresponds to the force-balance condition,
and the next $(n=2)$ to the torque-free condition. The following term $(n=3)$
ensures that the optimized boundary condition agrees with the measured
photospheric data, and that the last term $(n=4)$ controls the smoothing. The
2D-Laplace operator is designated by $\Delta$.

The aim of our preprocessing procedure is to minimize $L$ so that all terms
$L_{n}$, if possible, become small simultaneously. This will yield a surface
magnetic field:
\begin{equation}
{\vec{B}}_{min}=argmin(L)\label{thirty}
\end{equation}
Besides a dependence on the observed magnetogram, the solution in Eq.(\ref{twenty_five}) now also
depends on the coefficients $\mu_{n}$. These coefficients are a formal necessity because the terms $L_{n}$ represent
different quantities. By means of these coefficients, however, we can also give more or less weight to the individual
terms in the case where a reduction in one term opposes a reduction in another. This competition obviously exists
between the observation term $(n=3)$ and the smoothing term $(n=4)$.
The smoothing is performed consistently for all three magnetic field components.

To obtain Eq.(\ref{thirty}) by iteration, we need the derivative of $L$ with respect to each of the three field components at
 every node $(q)$ of the bottom boundary grid. We have, however, taken into account that $B_{r}$ is measured
with much higher accuracy than $B_{\theta}$ and $B_{\phi}$. This is achieved by assuming that the vertical component is invariable
compared to horizontal components in all terms where mixed products of the vertical and horizontal
field components occur, e.g., within the constraints \citep{Wiegelmann06sak}.
The relevant functional derivatives of $L$ are therefore\footnote{See Appendix B for partial derivative of
$L_{4}$ with respect to each of the three field components.}
\begin{equation}\begin{split}
\frac{\partial L}{\partial(B_{\theta})_{q}}=& 2\mu_{1}(B_{\theta}\sin^{2}\theta\cos\phi-
 B_{r}\sin\theta\cos\theta\cos\phi )_{q}\times
\\&\sum_{p}\big[E_{B}^{-}\sin\theta\cos\phi-B_{r}B_{1}\big]\sin\theta
\\&+2\mu_{1}(B_{\theta}\sin^{2}\theta\sin\phi- B_{r}\sin\theta\cos\theta\sin\phi )_{q}\times
\\&\sum_{p}\big[E_{B}^{-}\sin\theta\sin\phi-B_{r}B_{2}\big]\sin\theta
 \\&+2\mu_{1}( B_{\theta}\sin\theta\cos\theta+B_{r}\sin^{2}\theta)_{q}\times
 \\&\sum_{p}\big[E_{B}^{-}\cos\theta+B_{r}B_{\theta}\sin\theta\big]\sin\theta
\\ &+2\mu_{2}\Big[(B_{r}\sin\theta\sin\phi)_{q}\sum_{p} B_{r}B_{3}\sin\theta
\\&-(B_{r}\sin\theta\cos\phi)_{q}\sum_{p} B_{r}B_{4}\sin\theta\Big]
\\ &+2\mu_{3}(B_{\theta}-B_{\theta obs})_{q}+2\mu_{4}(\Delta(\Delta B_{\theta}))_{q},\label{thirty_one}
\end{split}\end{equation}
\begin{equation}\begin{split}
\\ \frac{\partial L}{\partial(B_{\phi})_{q}}=&2\mu_{1}(B_{\phi}\sin^{2}\theta\cos\phi+B_{r}\sin\theta\sin\phi )_{q}\times
\\&\sum_{p}\big[E_{B}^{-}\sin\theta\cos\phi-B_{r}B_{1}\big]\sin\theta
\\ &+2\mu_{1}(B_{\phi}\sin^{2}\theta\sin\phi- B_{r}\sin\theta\cos\phi )_{q}\times
\\&\sum_{p}\big[E_{B}^{-}\sin\theta\sin\phi-B_{r}B_{2}\big]\sin\theta
 \\ &+2\mu_{1}( B_{\phi}\sin\theta\cos\theta)_{q}\sum_{p}\big[E_{B}^{-}\cos\theta
 +B_{r}B_{\theta}\sin\theta\big]\sin\theta
\\ &+2\mu_{2}\Big[(B_{r}\cos\theta\cos\phi\sin\theta)_{q}\sum_{p}B_{r}B_{3}\sin\theta
\\ &+(B_{r}\cos\theta\sin\phi\sin\theta)_{q}\sum_{p}B_{r}B_{4}\sin\theta
\\ &+(B_{r}\sin^{2}\theta)_{q}\sum_{p}B_{r}B_{\phi}\sin^{2}\theta\Big]
+2\mu_{3}(B_{\phi}-B_{\phi obs})_{q}\\ &+2\mu_{4}(\Delta(\Delta B_{\phi}))_{q},\label{thirty_two}
\end{split}
\end{equation}
\begin{equation}
\frac{\partial L}{\partial(B_{r})_{q}}=2\mu_{3}(B_{r}-B_{r obs})_{q}+2\mu_{4}(\Delta(\Delta B_{r}))_{q}
\label{thirty_three}
\end{equation}
The optimization is performed iteratively by a simple Newton or Landweber iteration, which
replaces
\begin{equation}
(B_{r})_{q}\longleftarrow (B_{r})_{q}-\mu \frac{\partial L}{\partial(B_{r})_{q}},\label{thirty_four}
\end{equation}
\begin{equation}
(B_{\theta})_{q}\longleftarrow (B_{\theta})_{q}-\mu \frac{\partial L}{\partial(B_{\theta})_{q}},\label{thirty_five}
\end{equation}
\begin{equation}
(B_{\phi})_{q}\longleftarrow (B_{\phi})_{q}-\mu \frac{\partial L}{\partial(B_{\phi})_{q}},\label{thirty_six}
\end{equation}
at every step. The convergence of this scheme towards a solution of Eq.(\ref{twenty_five}) is
obvious: $L$ has to decrease monotonically at every step as long as Eqs.(\ref{twenty_six})-(\ref{twenty_eight})
have a nonzero component. These terms, however, vanish only if an extremum of $L$
is reached. Since $L$ is fourth order in $B$, this may not necessarily be a global minimum; in rare cases, if the step size
is handled carelessly, it may even be a local maximum. In practical calculation, this should not, however, be a problem
and from our experience we rapidly obtain a minimum ${\vec{B}}_{min}$ of $L$, once the parameters $\mu_{n}$ are specified
\citep{Wiegelmann06sak}.
\begin{figure*}[htp!]
   \centering
   \mbox{
       \includegraphics[bb=91 105 470 355,clip,height=4.5cm,width=6.0cm]{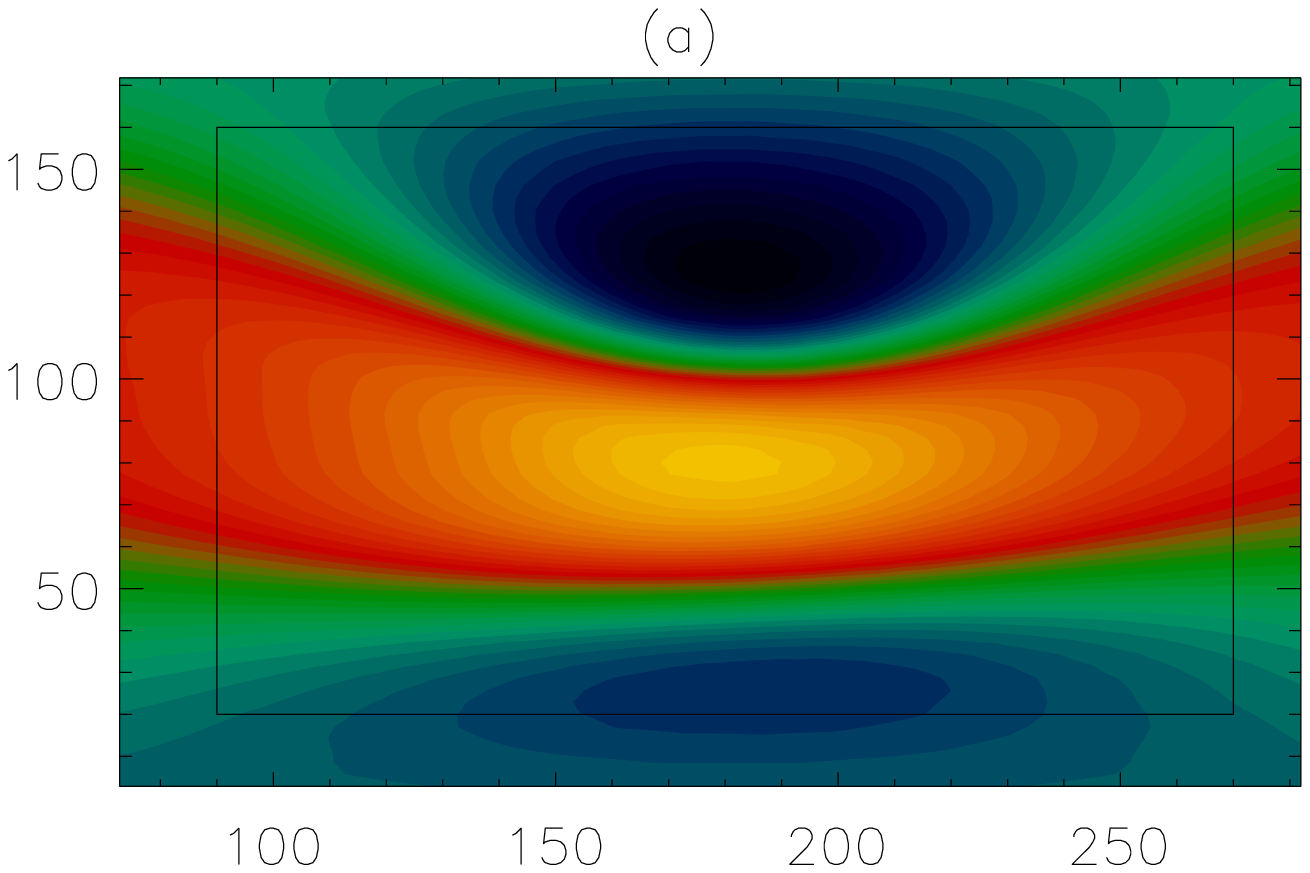}
     \includegraphics[bb=91 105 470 355,clip,height=4.5cm,width=6.0cm]{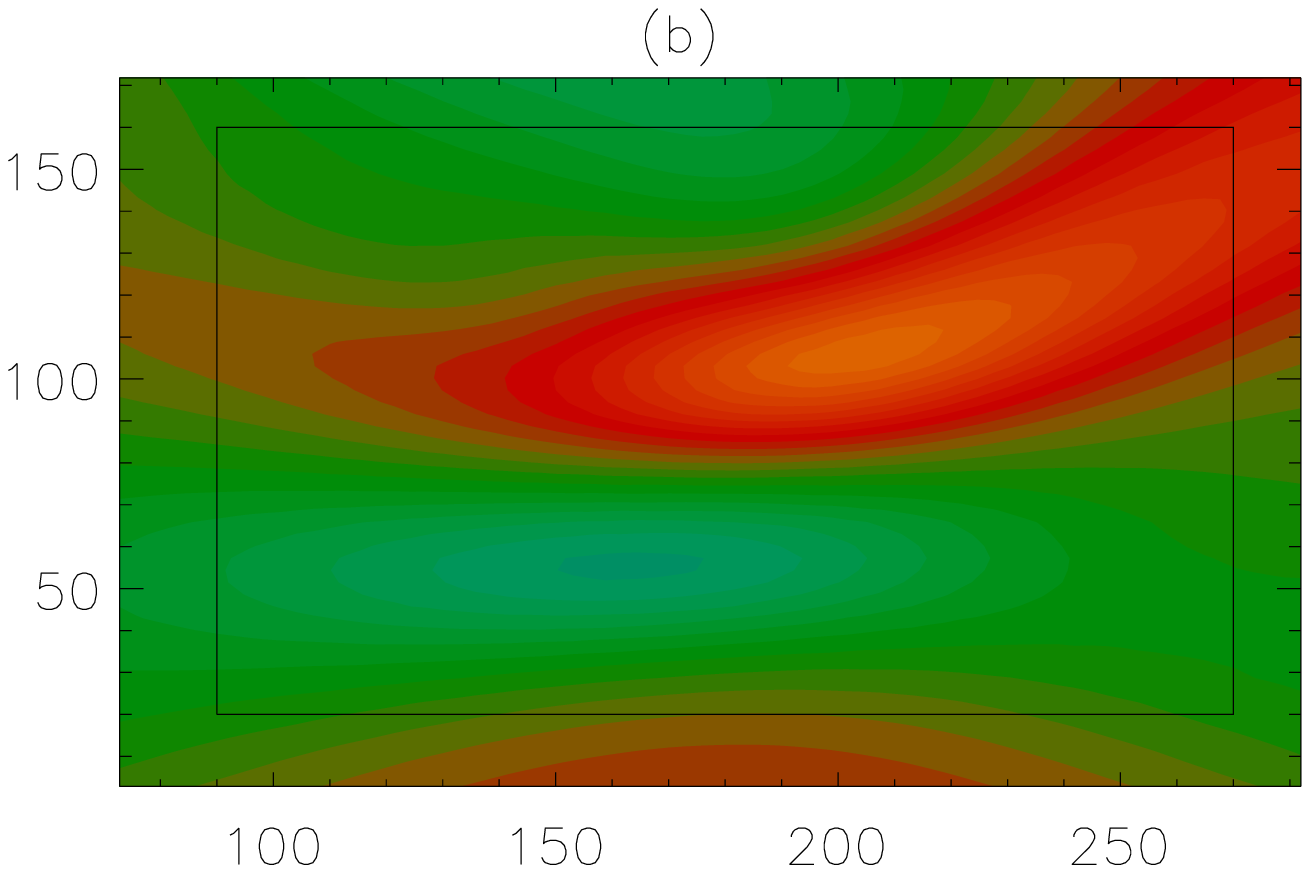}
    \includegraphics[bb=91 105 470 355,clip,height=4.5cm,width=6.0cm]{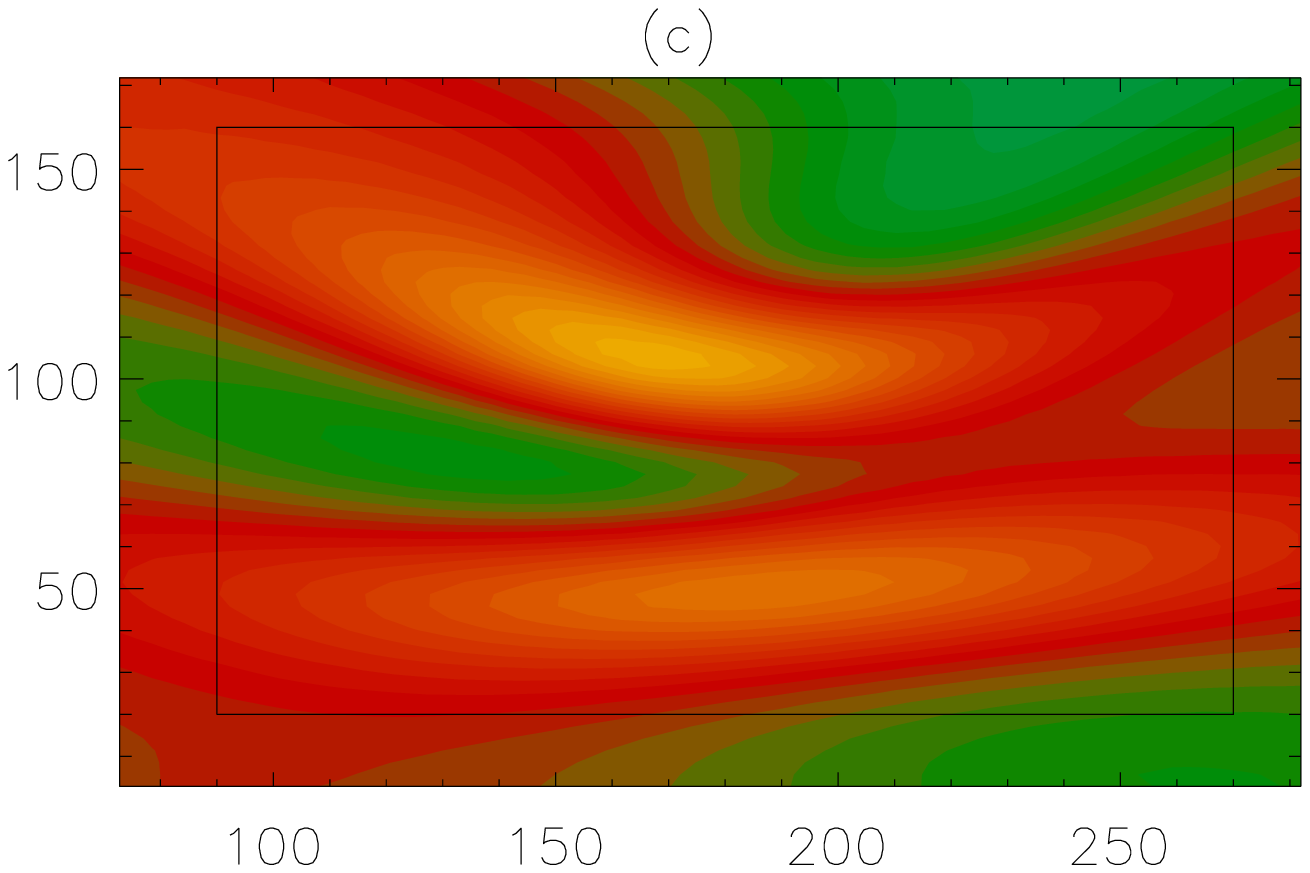}}
   \mbox{
   \includegraphics[bb=91 105 470 355,clip,height=4.5cm,width=6.0cm]{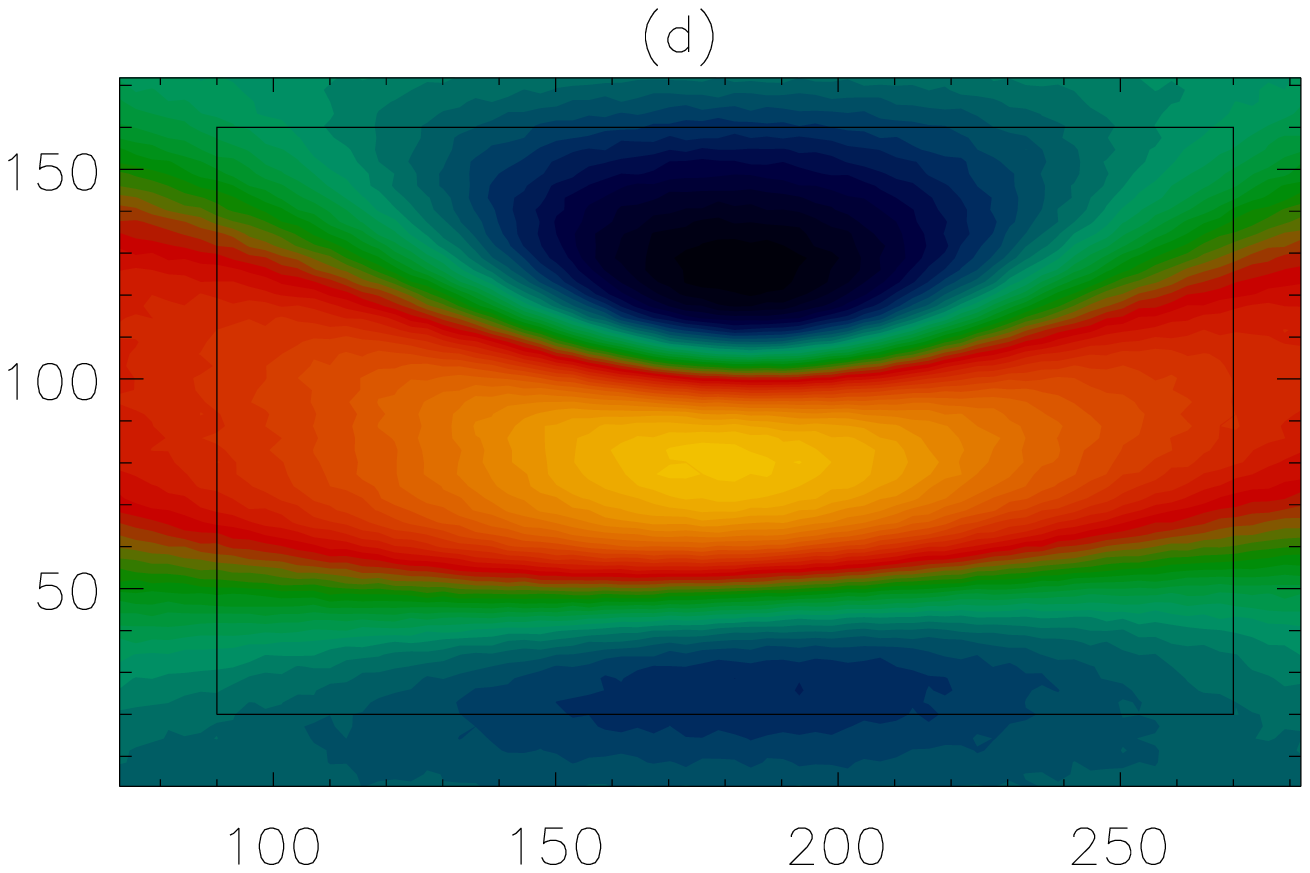}
   \includegraphics[bb=91 105 470 355,clip,height=4.5cm,width=6.0cm]{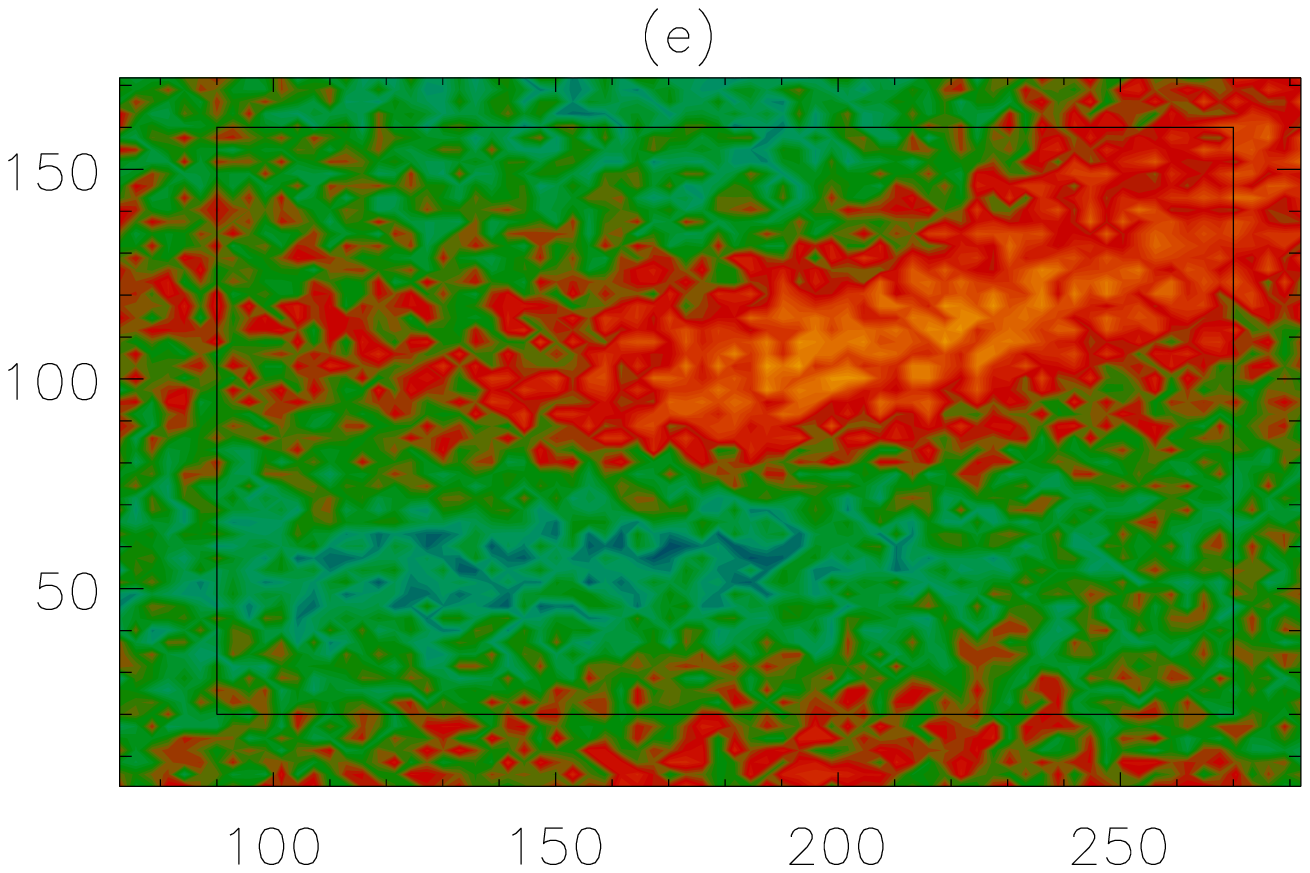}
   \includegraphics[bb=91 105 470 355,clip,height=4.5cm,width=6.0cm]{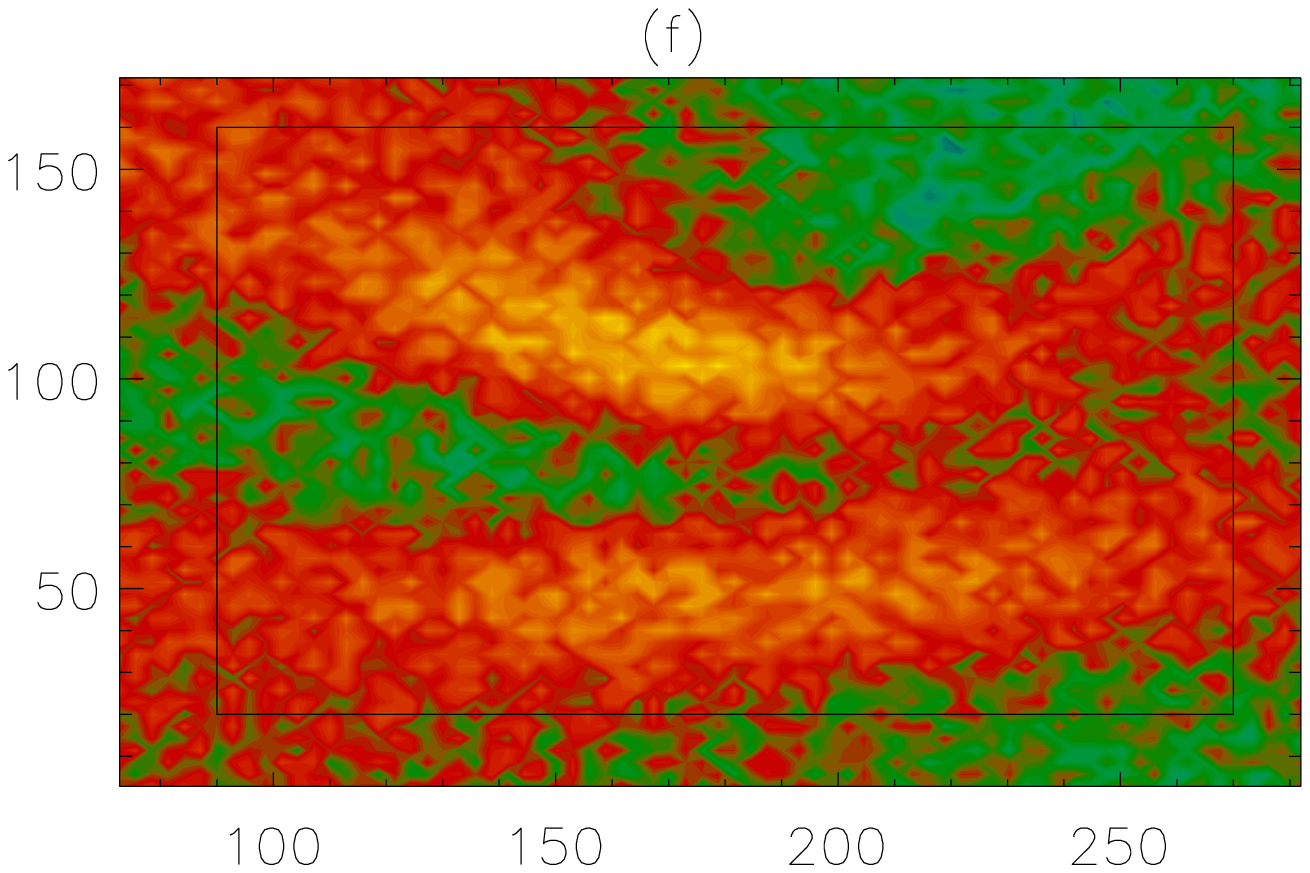}}
   \mbox{
   \includegraphics[bb=91 105 470 355,clip,height=4.5cm,width=6.0cm]{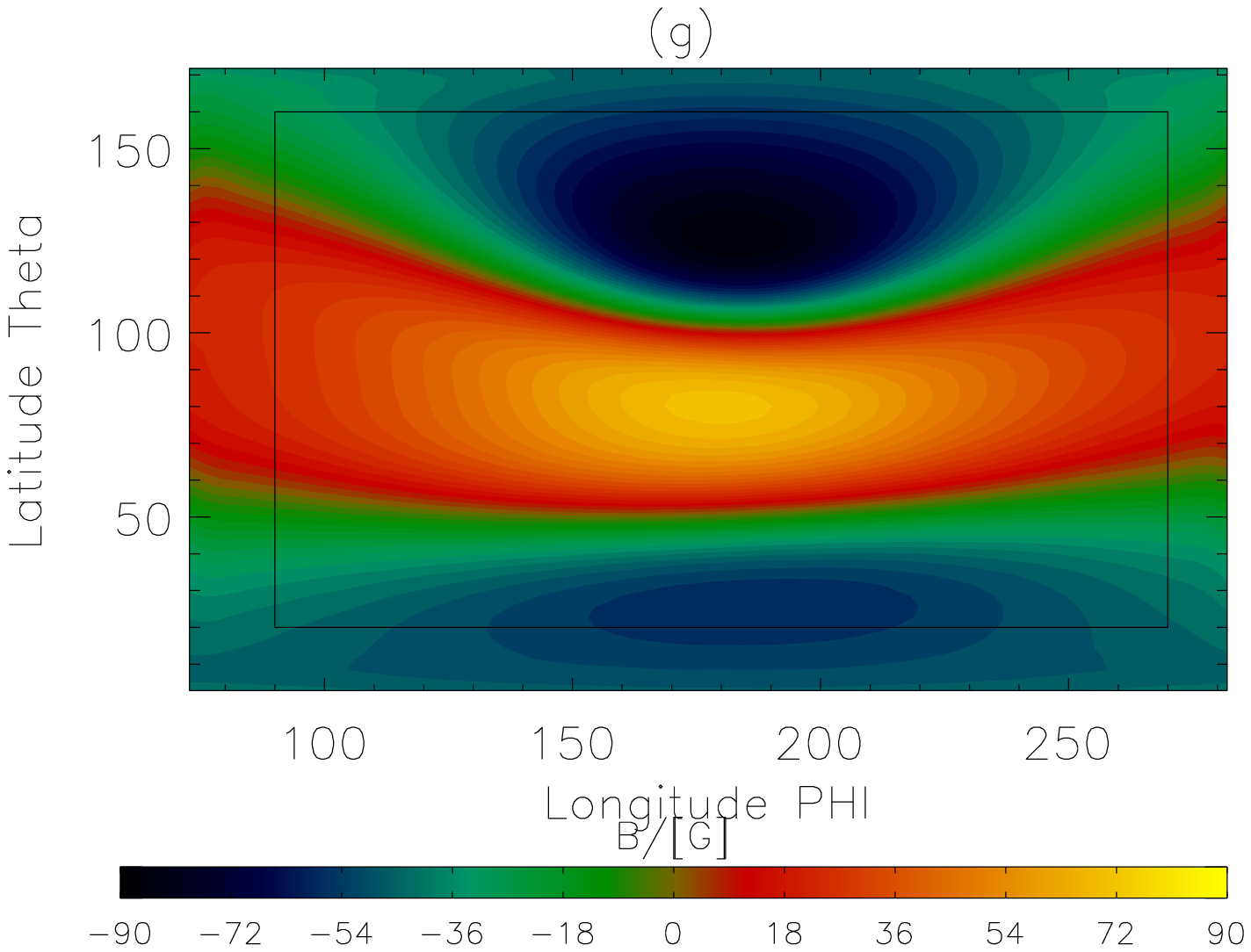}
   \includegraphics[bb=91 105 470 355,clip,height=4.5cm,width=6.0cm]{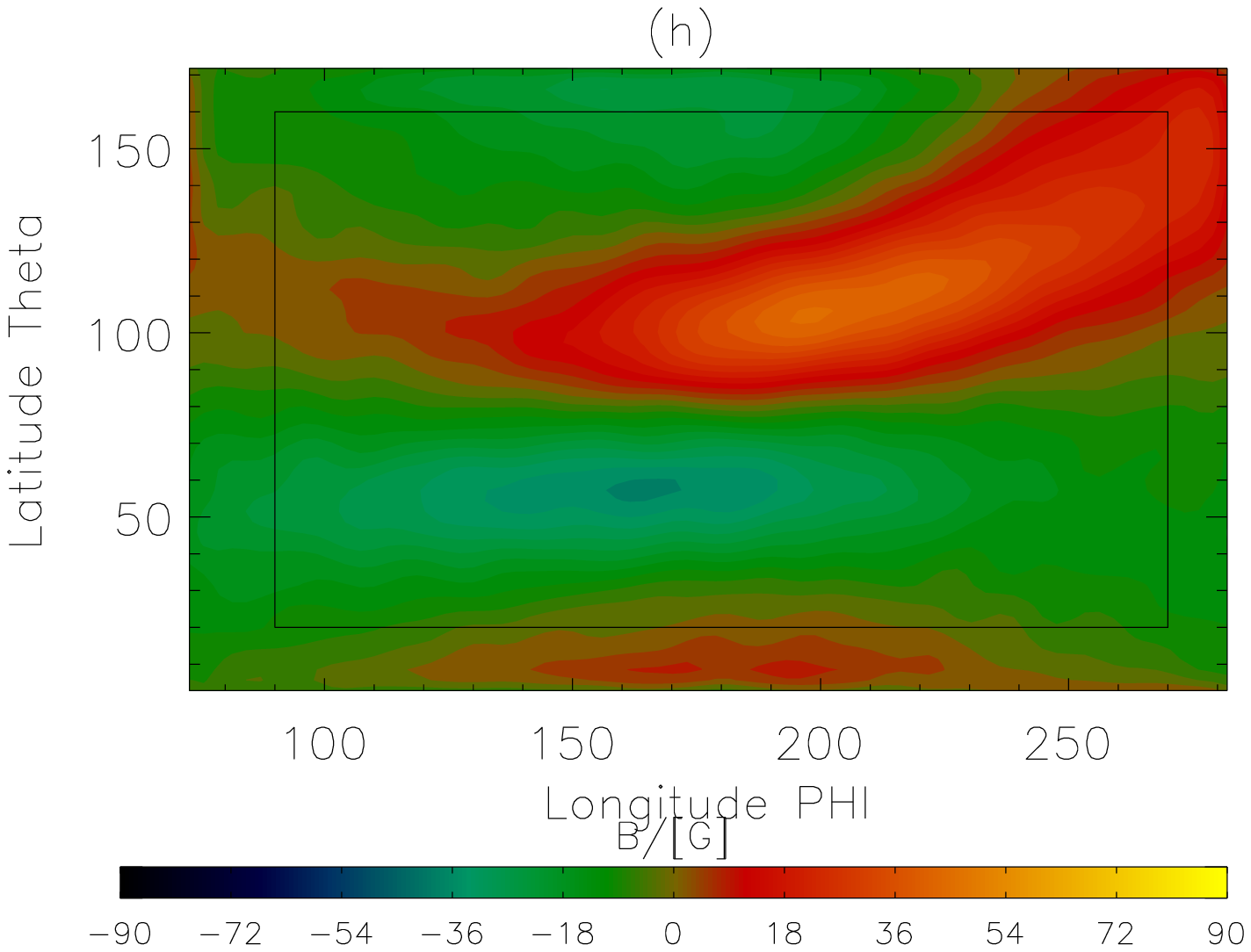}
   \includegraphics[bb=91 105 470 355,clip,height=4.5cm,width=6.0cm]{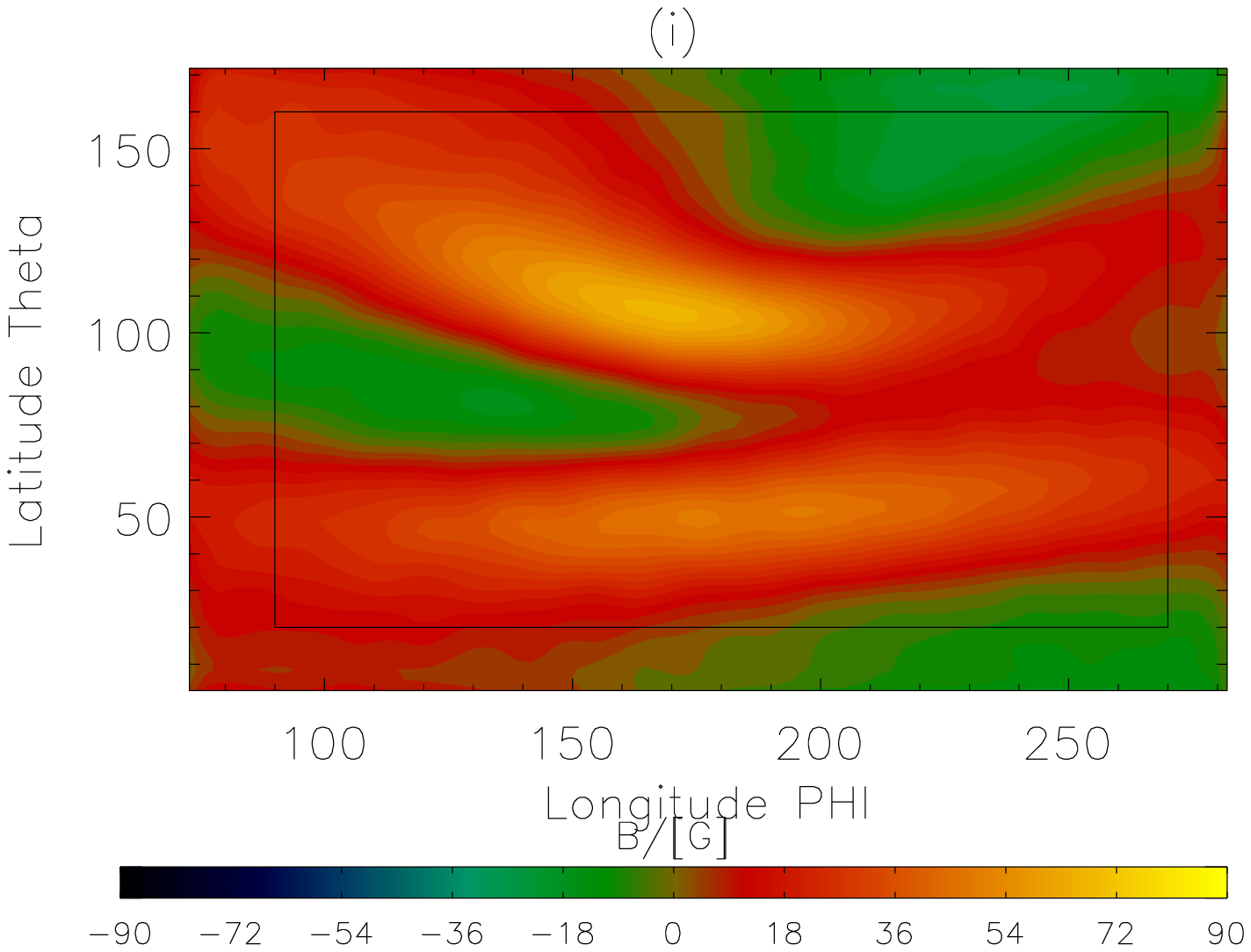}}
\includegraphics[bb=91 30 470 45,clip,height=0.3cm,width=14.0cm]{12529f3g.eps}
\includegraphics[bb=91 45 470 74,clip,height=0.8cm,width=14.0cm]{12529f3g.eps}
   \caption{ {\bf{Top row:}} vector magnetogram derive from the Low and Lou solution. From left to right the three components
$B_{r}$, $B_{\theta}$ \& $B_{\phi}$  are shown). {\bf{Middle row:}} the same magnetogram as in
the first row, but with noise added (noise model I). {\bf{Bottom row:}} magnetogram resulting from preprocessing
of the disturbed magnetogram shown in the second row. The magnetic fields are measured in gauss. The vertical and horizontal
axes show latitude, $\theta$ and longitude, $\phi$  on the photosphere respectively.}
\label{fig3}
\end{figure*}
\begin{figure*}[htp!]
   \centering
  \mbox{\subfigure[Original reference field]
   {\includegraphics[bb=120 60 445 350,clip,height=6.0cm,width=7.0cm]{12529f1a.eps}}
\subfigure[Potential field]
   {\includegraphics[bb=120 60 445 350,clip,height=6.0cm,width=7.0cm]{12529f1b.eps}}}
\mbox{\subfigure[Field from noisy data]
   {\includegraphics[bb=120 60 445 350,clip,height=6.0cm,width=7.0cm]{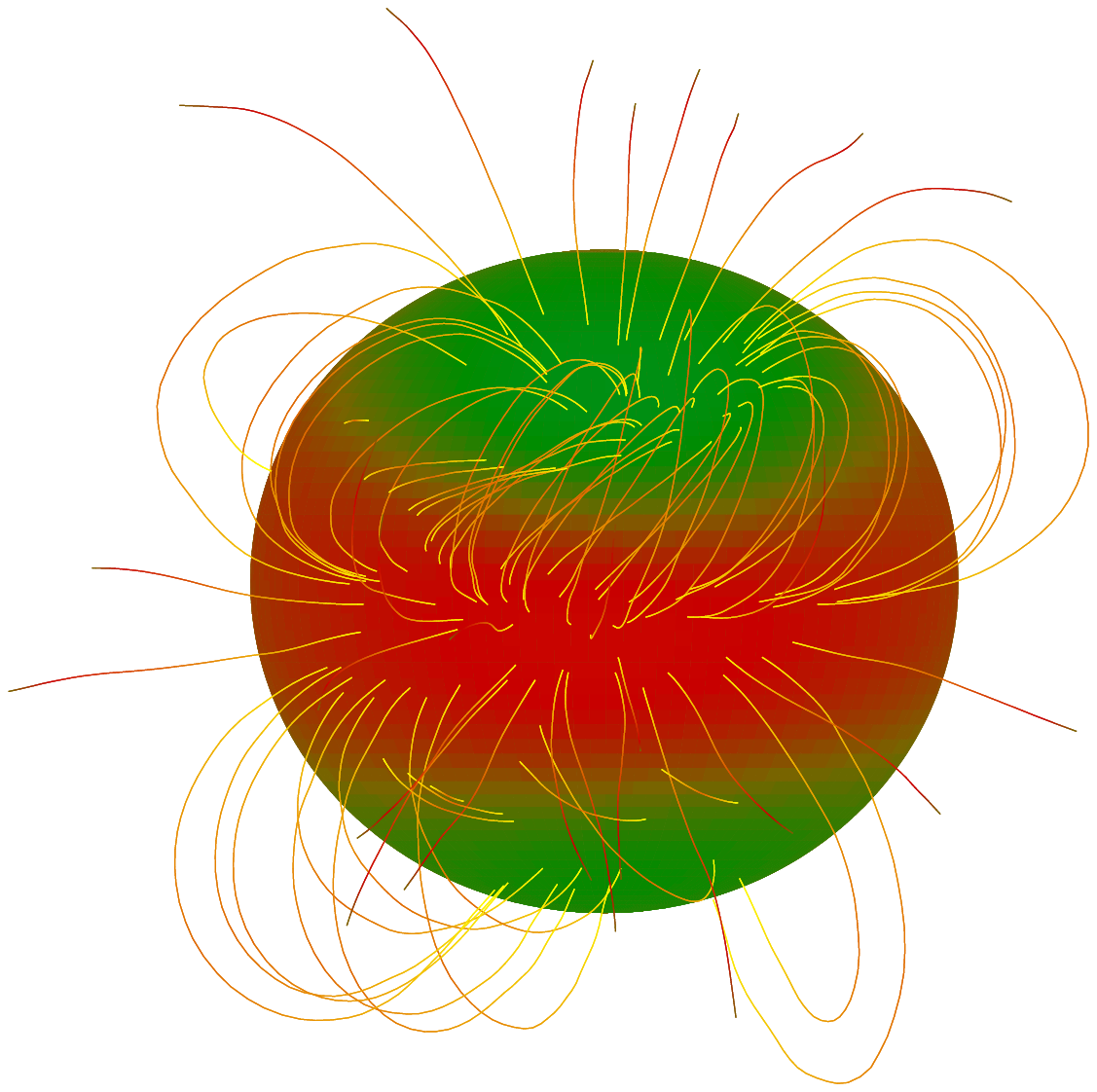}}
\subfigure[Field from preprocessed data]
   {\includegraphics[bb=120 60 445 350,clip,height=6.0cm,width=7.0cm]{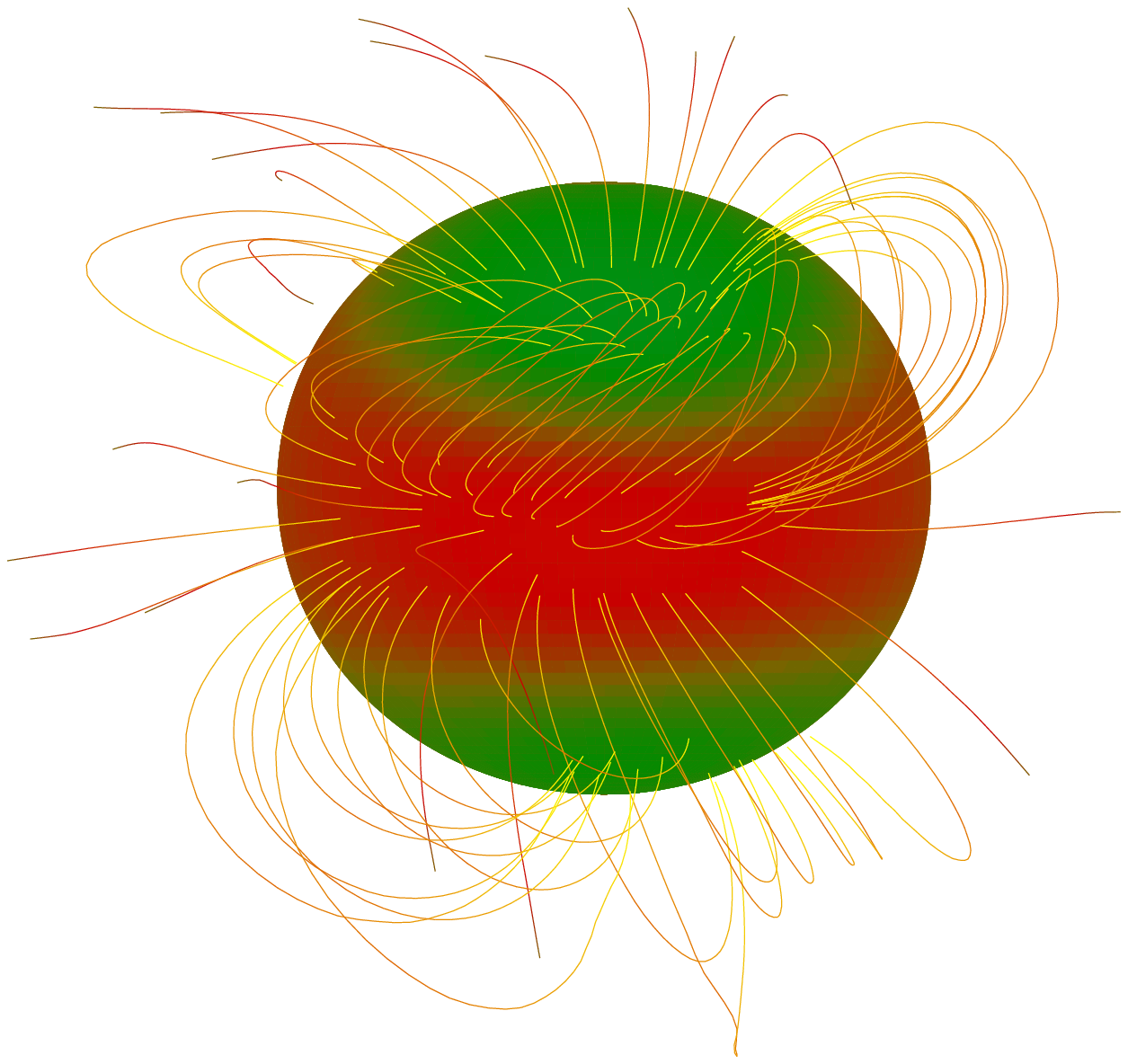}}}
     \caption{a) Some field lines for the original Low and Lou solution. b) Potential
field reconstruction. c) Nonlinear force-free reconstruction from noisy data (noise
model I) without preprocessing. d) Nonlinear force-free reconstruction from noisy
data (noise model I) after preprocessing the vector magnetogram with our newly
developed spherical code. }
\label{fig4}
\end{figure*}
\section{Application to different noise-models}
We extract the bottom boundary of the Low and Lou equilibrium and use it as input for our extrapolation code
\citep[see][]{Wiegelmann04}. This artificial vector magnetogram (see first row of Fig. \ref{fig3}) extrapolated
from a semi-analytical solution is of course in perfect agreement with the assumption of a force-free field
above (Aly-criteria) and the result of our extrapolation code was in reasonable agreement with the original.
True measured vector magnetograms are not ideal (and smooth) of course, and we simulate this effect by adding noise
to the Low and Lou magnetogram \citep{Wiegelmann06sak}. We add noise to this ideal solution in the
form:
\newline\indent\textbf{Noise model I:}\\
$\delta B_{i}=n_{l}\cdot r_{n}\cdot \sqrt{B_{i}},$ where $n_{l}$ is the noise level and $r_{n}$ a random number
in the range $-1....1$. The noise level was chosen to be $n_{l}=10.0$ for the transverse magnetic field
$(B_{\theta},B_{\phi})$ and $n_{l}=0.5$ for $B_{r}$. This mimics a real magnetogram (see the middle row of Fig.
\ref{fig3}) with Gaussian noise and significantly higher noise in the transverse components of the magnetic field.
\newline\indent\textbf{Noise model II:}\\
$\delta B_{i}=n_{l}\cdot r_{n},$ where $n_{l}$ is the noise level and $r_{n}$ a random number in the range $-1....1$.
The noise level was chosen to be $n_{l}=20.0$ for the transverse magnetic field $(B_{\theta},B_{\phi})$ and $n_{l}=1.0$ for
$B_{r})$. This noise model adds noise, independent of the local magnetic field strength.
\newline\indent\textbf{Noise model III:}\\
$\delta B_{r}=\textrm{constant}$, $\delta B_{t}=\frac{\delta
B^{2}_{tmin}}{\sqrt{B^{2}_{t}+B^{2}_{tmin}}}$, where we choose a constant noise level $\delta B_{r}$ of $1$ and a minimum
detection level $\delta B_{tmin}=20$. This noise model mimics the effect in which the transverse noise
level is higher in regions of low magnetic field strength \citep{Wiegelmann06sak}.
\begin{table*}
\caption {
 Figures of merit for the three different noise models with and without preprocessing along
 with model reference field and potential field.
 }
\hspace*{-0.5cm}
\centering
\begin{tabular}{lc|lllcc|rrrrrr}     
\hline \hline Model&Preprocessed&$L$&$L_1$&$L_2$&
$\parallel \nabla \cdot {\bf B} \parallel_{\infty}$&
$\parallel {\bf j } \times {\bf B} \parallel_{\infty} $ &
$C_{\rm vec}$&$C_{\rm CS}$&$E_{N}$&$E_{M}$&$\epsilon$ &Steps\\
\hline
Original &&$0.029$&$0.015$&$0.014$&$1.180$&$1.355$& $ 1$&$ 1$&$ 0$&$ 0$&$ 1$&$ $ \\
Potential &&$0.020$&$0.007$&$0.014$&$1.706$&$1.091$&$0.736$&$0.688$&$0.573$&$0.535$&$0.676$&$$ \\
Noise model I &No&$22.015$&$8.612$&$13.403$&$25.531$&$11.671$&$0.819$&$0.767$&$0.337$&$0.421$&$0.861$&$1337$ \\
Noise model I &Yes&$0.105$&$0.066$&$0.039$&$1.746$&$1.806$&$0.951$&$0.947$&$0.197$&$0.105$&$0.964$&$12191$ \\
Noise model II &No&$18.957$&$7.915$&$11.042$&$23.089$&$9.871$&$0.828$&$0.774$&$0.321$&$0.417$&$0.869$&$1484$ \\
 Noise model II &Yes&$0.097$&$0.057$&$0.040$&$1.533$&$1.617$&$0.963$&$0.951$&$0.191$&$0.099$&$0.971$&$11423$ \\
  Noise model III &No&$17.718$&$7.615$&$10.103$&$20.763$&$8.992$&$0.859$&$0.781$&$0.310$&$0.402$&$0.873$&$1497$ \\
  Noise model III &Yes&$0.081$&$0.043$&$0.038$&$1.382$&$1.407$&$0.979$&$0.957$&$0.189$&$0.098$&$0.982$&$10378$ \\
\hline
\end{tabular}
\end{table*}

The bottom row of Fig. \ref{fig3} shows the preprocessed vector magnetogram (for noise model I) after applying our procedure.
The aim of the preprocessing is to use the resulting magnetogram as input for a nonlinear force-free magnetic field extrapolation.

Figure \ref{fig4} shows in panel a) the original Low and Lou solution and in panel b) a corresponding
potential field reconstruction. In Fig. \ref{fig4} we present only the
inner region of the whole magnetogram (marked with black rectangular box in Fig.\ref{fig3}(a)) because the surrounding magnetogram
is used as a boundary layer (6 grid points) for our nonlinear force-free code. The computation was done on a $26 \times 60 \times 74$
 grid including a 6 pixel boundary layer towards the lateral and top boundary of the computational box $V$. In the remaining
 panels of Fig. \ref{fig4}, we demonstrate the effect of the  noise model (I) on the reconstruction. The noise
levels were chosen so that the mean noise was similar for all three noise models.  Fig. \ref{fig4} (c) shows a nonlinear
force-free reconstruction with noisy data (noise model I, magnetogram shown in the central panel of Fig.
\ref{fig3}), and Fig. \ref{fig4} (d) presents a nonlinear force-free reconstruction after preprocessing (magnetogram shown
in the bottom panel of Fig. \ref{fig3}). After preprocessing(see Fig. \ref{fig4} d), we achieve far closer agreement with
the original solution (Fig. \ref{fig4} a).  Field lines are plotted from the same photospheric footpoints in the positive polarity.

 For the other noise models II and III, we find that the preprocessed data agree more closely with the original Fig. \ref{fig4} (a).
 We check the correlation of the original solution with our reconstruction with help of the vector correlation function as defined
in (\ref{nine}).

\begin{figure}
   \centering
    \includegraphics[bb=90 45 496 342,clip,height=6.5cm,width=8.5cm]{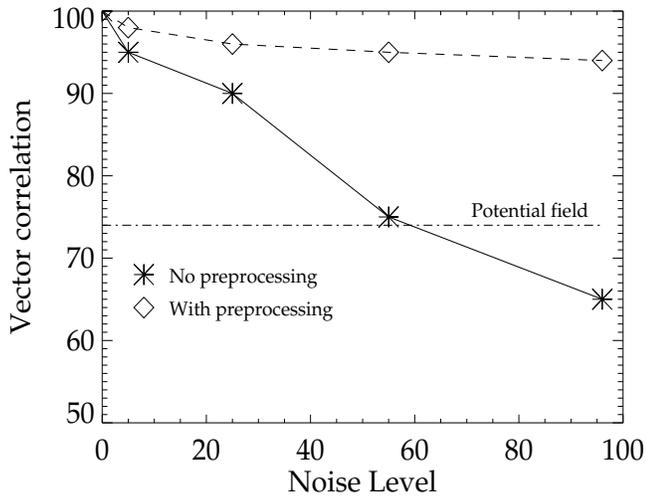}
      \caption{Vector correlation plotted against noise level for noise model I.}
      \label{fig5}
 \end{figure}

 Table $2$ confirms the visual inspection of Fig. \ref{fig4}. The correlation of the reconstructed magnetic field with the original
  improves significantly after preprocessing of the data for all noise models.
We knew already from previous studies \citep{Wiegelmann03,Wiegelmann04} that noise and inconsistencies in vector magnetograms
have a negative influence on the nonlinear force-free reconstruction, and the preprocessing routine described in this paper shows us
how to overcome these difficulties in the case of spherical geometry. As indicated by Fig. \ref{fig5}, the higher the noise level
we have added to the original magnetogram, the smaller the vector correlation will be for the field reconstructed from the
magnetogram with noise, compared with the reference field. However, the corresponding vector correlations for the field
reconstructed from the preprocessed magnetogram has no significant change as the code largely removes the
noise we have added to the original magnetogram with different noise levels.
\section{Conclusion and outlook}
In this paper, we have developed and tested the optimization method for the reconstruction of nonlinear
force-free coronal magnetic fields in spherical geometry by restricting the code to limited parts of the Sun, as
suggested by \citet{Wiegelmann07}.
 The optimization method minimizes a functional consisting of a quadratic form of the force
balance and the solenoidal condition. Without a weighting function, all the six boundaries are equal
likely to influence the solution. The effect of top and lateral boundaries can be reduced by introducing a boundary layer
around the physical domain \citep{Wiegelmann04}. The physical domain is a wedge-shaped area within which we
 reconstruct the coronal magnetic field that is consistent with the photospheric vector magnetogram data.
 The boundary layer replaces the hard lateral and top boundary used previously. In the physical domain, the weighting function
is unity. It drops monotonically in the boundary layer and reaches zero at the boundary of the computational box.
At the boundary of the computational box, we set the field to have the value of the potential field computed from $B_{r}$ at
the bottom boundary. Our test calculations show that a finite-sized weighted boundary yields far more reliable results.
The depth $nd$ of this buffer boundary influences the quality of reconstruction, since the magnetic
flux in these test cases is not concentrated well inside the interior of the box.

In this work, we have presented a method for preprocessing vector magnetogram data to be able to use the preprocessing result
 as input for a nonlinear force-free magnetic field extrapolation with help of an optimization code in spherical geometry.
We extended the preprocessing routine developed by \citet{Wiegelmann06sak} to spherical geometry. As a first test of the
method, we use the Low and Lou solution with added noise from different noise models.
A direct use of the noisy photospheric data for a nonlinear force-free extrapolation showed no good agreement with the
original Low and Lou solution, but after applying our newly developed preprocessing method
we obtained a reasonable agreement with the original. The preprocessing method changes the boundary data within their noise limits
to drive the magnetogram towards boundary conditions that are consistent with the assumption of a force-free field above.
The transverse field components with higher noise level are modified more than the radial components.

To carry out the preprocessing, we use a minimization principle.
On the one hand, we control the final boundary data to be as close as possible (within the noise level)
to the original measured data, and the data are forced to fulfill the consistency criteria and be sufficiently smooth.
Smoothness of the boundary data is required by the nonlinear force-free extrapolation code, but also necessary physically because
the magnetic field at the basis of the corona should be smoother than in the photosphere, where it is measured. In addition to
these, we found that adding a larger amount of noise to the magnetogram decreases its vector correlation with the model reference
 field whenever we reconstruct it without preprocessing.

We plan to use this newly developed code for future missions such as SDO (Solar Dynamic Observatory)
when full disc magnetogram data become available.
\begin{acknowledgements}Tilaye Tadesse acknowledges a fellowship of the International Max-Planck Research School
at the Max-Planck Institute for Solar System Research and the work of T.
Wiegelmann was supported by DLR-grant $50$ OC $453$  $0501$. The authors would like to thank referee for
his/her constructive and helpful comments.
\end{acknowledgements}
\bibliographystyle{aa}
\bibliography{12529_bibtex}
\newpage
\begin{appendix}
\section{Torque-balance equations}
To solve the torque-balance equations in Eqs.(\ref{twenty_two})-(\ref{twenty_four}), we
assume that the volume integral of torque in the computational box vanishes to fulfill the force-free
criteria.
\begin{equation}
     \int_{V}\big({\vec{ r}}\times {\vec{F}}\big)dV=0,\label{A1}
\end{equation}
where the force $\vec{F}$ is
\begin{equation}
     \vec{F}=\frac{1}{4\pi}(\nabla\times \vec{B})\times\vec{B}\label{A2}
\end{equation}
Substituting Eq.(\ref{A2}) into Eq.(\ref{A1}) and using vector identity
\begin{equation}
    {\vec{\nabla}}({\vec{A}}\cdot {\vec{ B}})={\vec{ A}}\times ( {\vec{\nabla}}\times{\vec{B}}) + {\vec{B}}\times
    ({\vec{\nabla}} \times{\vec{ A}})+({\vec{A}}\cdot{\vec{\nabla}}){\vec{ B}}+({\vec{ B}}\cdot{\vec{\nabla}}){\vec{ A}}
     ,\label{A3}
\end{equation}
along with the Gauss divergence theorem we can find the following expression
\begin{equation}
   \int_{V}\big({\vec{ r}}\times {\vec{F}}\big)dV=\frac{1}{2}\int_{S}B^{2}\big({\vec{ r}}\times {d\vec{ S}}\big)-
   \int_{S}\big({\vec{r}}\times {\vec{B}}\big)\big({\vec{B}}\cdot {d\vec{S}}\big)=0.\label{A4}
\end{equation}
The vector $d\vec{S}$ is directed into the volume $V$, with $d\vec{S}=R^{2}_{\sun}\sin\theta d\theta d\phi
\hat{{\vec{e}}}_{r}$. The origin of the vector $\vec{r}=R_{\sun}\hat{{\vec{e}}}_{r}$ is taken to be the centre of the Sun.
Therefore, Eq.(\ref{A4}) reduces to the form
\begin{equation}
      \int_{S}\big({\vec{r}}\times {\vec{B}}\big)\big({\vec{B}}\cdot {d\vec{S}}\big)=0,\label{A5}
\end{equation}
   since $\big({\vec{ r}}\times {d\vec{ S}}\big)=0$.  Where
$\vec{B}=B_{r}\hat{\vec{e}}_{r}+B_{\theta}\hat{\vec{e}}_{\theta}+B_{\phi}\hat{\vec{e}}_{\phi}$, we
have $\vec{B}\cdot d\vec{S}=R^{2}_{\sun}B_{r}\sin\theta d\theta
d\phi=R^{2}_{\sun}B_{r}d\Omega$. By substituting the vector $\vec{r}\times\vec{B}$ into Eq.(\ref{A5}), one can find
\begin{equation}
\int_{S}\Big[-R_{\sun}B_{\phi}\hat{\vec{e}}_{\theta}+R_{\sun}B_{\theta}\hat{\vec{e}}_{\phi}\Big]B_{r}R^{2}_{\sun}d\Omega=0.
\label{A6}
\end{equation}
Hence the torque balance along the $x$-component will be
\begin{equation}\hat{\vec{e}}_{x}\cdot\Big[\int_{S}\big(-R_{\sun}B_{\phi}\hat{\vec{e}}_{\theta}
+R_{\sun}B_{\theta}\hat{\vec{e}}_{\phi}\big) B_{r}R^{2}_{\sun}d\Omega\Big]=0,\label{A7}
\end{equation}
where $\hat{\vec{e}}_{x}$ is the unit vector along the $x$-axis. Normalizing this equation by assuming that $R_{\sun}=1$,
Eq.(\ref{A7}) reduces to
\begin{equation}
\int_{S}B_{r}\big(B_{\phi}\cos\theta\cos\phi+B_{\theta}\sin\phi\big)d\Omega=0,\label{A8}
\end{equation}
 Similarly one can obtain Eqs. (\ref{twenty_three}) and (\ref{twenty_four}).
\section{Partial derivative of $L_{4}$}
We derive the partial derivative of $L_{4}$ with respect to each of the three magnetic field
components in its discretized form as indicated in Eqs.(\ref{thirty_one})-(\ref{thirty_three}).
We used a five-point stencil on the photospheric boundary for Laplace in $L_{4}$. Those derivatives are carried out at
every node (q) of the bottom boundary grid. The partial derivative of Eq. (\ref{twenty_nine}) with respect to
$B_{r}$, for instance can be written as
\begin{equation}
     \frac{\partial L_{4}}{\partial (B_{r})_{q}}=2\sum_{p}(\Delta B_{r})_{p}\frac{\partial}{\partial (B_{r})_{q}}(\Delta B_{r})_{p}
      \label{B1}
\end{equation}
We demonstrated the effect of the derivative by using the conventional Laplacian $\Delta B_{r}$
in one dimension using three-point stencil with geometry-dependent coefficients $c$ \& $a$. Then
\begin{equation}
(\Delta B_{r})_{p}=a(B_{r})_{p-1}+c(B_{r})_{p}+a(B_{r})_{p+1}
      ,\label{B2}
\end{equation}
and after substituting Eq.(\ref{B2}) into the derivative term in Eq.(\ref{B1}), we find
\begin{equation}\begin{split}
\frac{\partial}{\partial (B_{r})_{q}}(\Delta B_{r})_{p}=&\frac{\partial}{\partial (B_{r})_{q}}\big(a(B_{r})_{p-1}+
c(B_{r})_{p}+a(B_{r})_{p+1}\big)\\
=&a\delta _{p-1,q}+c\delta _{p,q}+a\delta _{p+1,q}\label{B3}
\end{split}\end{equation}
Therefore, using equation Eq.(\ref{B3}), we can reduce Eq.(\ref{B1}) to
\begin{equation}\begin{split}
     \frac{\partial L_{4}}{\partial (B_{r})_{q}}=&2\sum_{p}(\Delta B_{r})_{p}\big(a\delta _{p-1,q}+
     c\delta _{p,q}+a\delta _{p+1,q}\big)\\
     =&2\sum_{p}\big[a(\Delta B_{r})_{q+1}+
     c(\Delta B_{r})_{q}+a(\Delta B_{r})_{q-1}\big]\\
     =&2\sum_{p}\big(\Delta(\Delta B_{r})\big)_{q}.\label{B4}
\end{split}\end{equation}
One can similarly derive the partial derivative of $L_{4}$ with
respect to the other two field components.
\end{appendix}
\end{document}